\documentclass{aa}  

\usepackage{graphicx}
\usepackage{txfonts}
\usepackage{lipsum}
\usepackage{lscape}
\usepackage{placeins}
\usepackage{subfig}
\usepackage{tabularx}
\usepackage{newtxtext,newtxmath}
\usepackage[T1]{fontenc}
\usepackage{eucal}
\usepackage{nicefrac}
\usepackage{amssymb}
\usepackage{latexsym}
\usepackage{amsmath, mathrsfs, bm}
\usepackage{eqnarray}
\usepackage{comment}
\usepackage{url}
\usepackage{cases}
\usepackage{epsfig}
\usepackage{color}
\usepackage{grffile}
\usepackage{float}
\usepackage{soul}
\usepackage{enumerate}
\usepackage{multirow}
\usepackage{threeparttable}
\usepackage{booktabs}
\usepackage{xspace}
\usepackage{bm}
\usepackage{times}
\usepackage{etoolbox}
\usepackage{multirow}
\usepackage{hyperref}
\hypersetup{hidelinks}

\usepackage{xcolor}
\definecolor{green}{RGB}{50,205,50}

\begin{document}

\title{Estimating the peak energy of \textit{Swift} gamma-ray bursts using supervised machine learning}

\titlerunning{Estimating the peak energy of \textit{Swift} GRBs using supervised machine learning}

\author{Wan-Peng Sun\inst{1}
        \and Si-Yuan Zhu\inst{1}
    \and Da-Ling Ma\inst{1}
    \and Fu-Wen Zhang\inst{1,2}\fnmsep\thanks{Corresponding authors: fwzhang@pmo.ac.cn}
    }

\institute{College of Physics and Electronic Information Engineering, Guilin University of Technology, Guilin 541004, China\\
         \and Key Laboratory of Low-dimensional Structural Physics and Application, Education Department of Guangxi Zhuang Autonomous Region, Guilin 541004, China}

\date{Received 4 January 2026 / Accepted 2 March 2026}

\abstract
{Gamma-ray bursts (GRBs) are among the most energetic explosive phenomena in the Universe, and their peak energy ($E_{\rm p}$) is a key physical quantity for understanding the prompt emission mechanism. However, due to the limited energy coverage of the \textit{Swift} satellite, a large fraction of \textit{Swift} GRBs lack reliable peak energy measurements.  Therefore, developing an accurate and efficient method for estimating $E_{\rm p}$ is of great importance. In this work, we propose a method based on the SuperLearner framework that integrates multiple supervised machine learning algorithms to estimate the $E_{\rm p}$ of \textit{Swift}/BAT GRBs. We used the \textit{Swift}/BAT observational data from December 2004 to September 2022 as training features, and adopted the peak energies of 516 GRBs jointly detected by \textit{Swift} and either \textit{Fermi}/GBM or Konus-Wind as training labels. After training and testing multiple supervised models, the final SuperLearner ensemble yields a more robust and reliable predictive model. In 100 iterations of five-fold cross-validation, the estimated $E'_{\rm p}$ values show a tight correlation with the observed $E_{\rm p}$, with an average Pearson correlation coefficient of $r = 0.72$. Compared with previous Bayesian estimates, our model provides estimations that are likely closer to the true values. Based on the trained model, we further estimated the peak energies of 650 \textit{Swift} GRBs, significantly increasing the number of GRBs with estimated peak energies and providing new statistical support for constraining GRB emission mechanisms and energy origins.}


\maketitle

\section{Introduction} \label{sec:intro}

Gamma-ray bursts (GRBs) are among the most energetic and active high-energy transients in the Universe, serving as key probes of extreme physical processes. Their prompt emission directly traces the initial energy dissipation of ultra-relativistic jets, providing insight into jet dynamics and radiation mechanisms. This emission typically exhibits a nonthermal spectrum, which in the $\nu F_{\nu}$ representation shows a prominent peak that corresponds to the peak energy ($E_{\rm p}$). As a fundamental physical quantity characterizing the spectral properties, $E_{\rm p}$ reflects the characteristic energy of the radiating particles, encodes information on radiation mechanisms and jet dynamics \citep{2001ApJ...555..540L,2011ApJ...732...49P,2004RvMP...76.1143P}, and plays a central role in spectrum-energy correlations. These include the $E_{\rm p,z}$-$E_{\rm iso}$ relation (known as the Amati relation), the $E_{\rm p,z}$-$L_{\rm iso}$ relation (Yonetoku relation), and other empirical correlations \citep{Amati2002, Amati2006, 2003MNRAS.345..743W, 2004ApJ...609..935Y, Ghirlanda:2008tu, 2015ApJ...813..116L}, where $E_{\rm p,z}$ is rest frame peak energy, $E_{\rm p,z}$=$E_{\rm p}(1+z)$, and $E_{\rm iso}$ and $L_{\rm iso}$ are the isotopic energy and peak luminosity of the GRB prompt emission, respectively. Such correlations establish GRBs as potential probes of high-redshift astrophysics and cosmology \citep{Han:2023exn,Du:2025csv,2026EPJC...86....8H}.

Accurately measuring $E_{\rm p}$ is challenging. Obtaining a reliable peak energy typically requires broadband spectral coverage -- from tens of keV to several MeV -- to capture the full spectral shape across both the low- and high-energy ranges. GRB spectra are commonly fitted with a broken power-law (PL) function, typically represented by the Band function \citep{1993ApJ...413..281B}, which consists of a low- and a high-energy component. In some cases, the spectra can also be described well by a cutoff power-law (CPL) model. For a subset of GRBs, the high-energy cutoff cannot be detected due to the limited energy range of the instruments, and their spectra are thus fitted with a simple PL model \citep{2008ApJS..175..179S, 2011ApJS..195....2S}, making it difficult to determine the peak energy.

Thanks to the rapid and accurate localization capabilities of the \textit{Swift} satellite, we now have a large sample of GRBs with measured redshifts, which has enabled a deeper exploration of their intrinsic properties and progenitors, including the evolution of their early afterglows. However, the \textit{Swift} Burst Alert Telescope (BAT) operates in a relatively narrow energy range of 15-150 keV, which is significantly lower than the typical GRB prompt emission energies (tens of keV to several MeV; \citealt{2014ApJS..211...12G, 2014A&A...561A..25B}). As a result, \textit{Swift} is often unable to provide reliable constraints on the spectral parameters during the prompt emission phase. Specifically, owing to the inherent limitations of the BAT detector, the spectra of a large fraction of \textit{Swift} GRBs can only be modeled using a PL function \citep{Zhang:2006uj, 2006HEAD....9.1704S}. In such cases, the observed photon index ($\Gamma$) merely reflects the spectral slope along the tail of the Band function rather than the intrinsic low-energy index ($\alpha$), thereby obscuring the spectral curvature around the peak energy and precluding a reliable determination of their peak energies. When \textit{Swift} GRBs are jointly detected by other broadband instruments such as Konus-Wind or \textit{Fermi} Gamma-ray Burst Monitor (GBM), their spectra can typically be fitted with the Band function; however, such joint detections account for only a small fraction of the \textit{Swift} sample. Therefore, obtaining reliable estimates of $E_{\rm p}$ for \textit{Swift} GRBs has long been a major challenge but is crucial for systematically investigating the intrinsic radiation mechanisms of GRBs.

Several studies have attempted to estimate $E_{\rm p}$ using empirical correlations involving the spectral index \citep{2012MNRAS.424.2821V, 2007ApJ...655L..25Z, 2006HEAD....9.1704S} or spectral hardness \citep{2005ChJAA...5..151C, Zhang:2006uj, 2010MNRAS.407.2075S}. These works typically combine observations from multiple instruments to construct more reliable estimation models. However, linear relations constructed from only two parameters often exhibit large intrinsic scatter and strong sample dependence. In contrast, \citet{2007ApJ...671..656B} introduced a Bayesian approach that incorporates additional observational parameters and prior knowledge, yielding more accurate estimates of $E_{\rm p}$. More recently, \citet{2024arXiv241208226L} exploited the spectral curvature within the BAT band to infer lower limits on $E_{\rm p}$ for bursts with $E_{\rm p}>150$ keV, providing a complementary pathway when direct constraints are unavailable.

However, traditional methods for estimating $E_{\rm p}$ are generally constrained by their reliance on crude, low-dimensional empirical correlations. To overcome this limitation, we seek methods capable of incorporating richer observational information and capturing the complex, nonlinear relationships between GRB observables and $E_{\rm p}$. With the rapid advancement of data-driven techniques, machine learning (ML) provides precisely such capabilities and has been increasingly applied in astrophysics \citep{WOS:000839971300001,2024A&C....4800851F,2025Univ...11..355W}, including for the classification of transient sources such as GRBs \citep{2024MNRAS.527.4272C,2024MNRAS.532.1434Z,2025MNRAS.541.3236Z,2025A&A...702A.173Z} and fast radio bursts (FRBs; \citealt{2023MNRAS.518.1629L,2025ApJ...980..185S,2026ApJ...998..339S}), as well as the estimation of GRB redshifts \citep{2024MNRAS.529.2676A,2025A&A...698A..92N}. ML thus offers a flexible framework for modeling such high-dimensional dependences. In this work, we focus on \textit{Swift} GRB observations and apply supervised ML models to estimate $E_{\rm p}$ from multiple observable features.

The ML framework used in this study includes several base learners -- random forest, extreme gradient boosting, linear regression, and kernel ridge regression -- which, after extensive training, exhibit good stability and reliable generalization. These models were then integrated using the SuperLearner ensemble method, which assigns optimal weights based on cross-validated performance. This strategy has proven highly effective for estimating the redshifts of active galactic nuclei and GRBs \citep{2021ApJ...920..118D, 2022ApJS..259...55N,2024ApJS..271...22D,2025ApJS..277...31D}, enabling the ensemble to exploit the complementary strengths of individual algorithms and achieve higher predictive accuracy.

In Sect. \ref{sec:data} we describe the sample and the data selection criteria used in this work. In Sect. \ref{sec:method} we present the selection of predictive parameters, the dataset partitioning strategy, and a detailed discussion of the training and optimization of the supervised ML algorithms. Section \ref{sec:resu} reports the main estimation results and compares the performance of the SuperLearner ensemble method with other approaches to estimating the peak energy of \textit{Swift} GRBs. Finally, Sect. \ref{sec:discu} summarizes our work.

\section{Data sample selection} \label{sec:data}
\subsection{Data} \label{ssec:data}
Since 2004, \textit{Swift}/BAT has continuously monitored the sky, providing a large dataset for GRB studies. In this work, we selected BAT observations from December 2004 to September 2022, retrieved from the \textit{Swift} archive\footnote{\url{https://swift.gsfc.nasa.gov/archive/}}, comprising a total of 1557 GRBs (hereafter referred to as the BAT sample). Among them, 229 GRBs can be well fit with the CPL model and have reliably determined $E_{\rm p}$ values; these GRBs were therefore excluded from our generalization sample. The remaining 1328 GRBs can only be fitted with a PL model and lack measured $E_{\rm p}$ values.

For training the ML models, we used $E_{\rm p}$ values from GRBs simultaneously observed by \textit{Swift} and the broadband instruments \textit{Fermi}/GBM and Konus-Wind. The \textit{Fermi}/GBM sample includes GRBs detected between June 2008 and September 2022 that are temporally coincident with \textit{Swift} triggers and have consistent sky positions (RA/Dec), with data obtained from the online \textit{Fermi} GRB catalog\footnote{\url{https://heasarc.gsfc.nasa.gov/W3Browse/fermi/fermigtrig.html}}. The Konus-Wind sample consists of GRBs observed concurrently with \textit{Swift} between December 2004 and September 2022, with data primarily drawn from \citet{2016ApJS..224...10S} and \citet{2017ApJ...850..161T,2021ApJ...908...83T}. For both \textit{Fermi} and Konus-Wind spectra, only GRBs whose spectra are best fitted by the Band or CPL functions were selected to ensure the reliability of $E_{\rm p}$ values.

\subsection{Feature selection and dataset partitioning} \label{ssec:Fselection}

Based on previous experience, to ensure that supervised ML algorithms achieve stable and reliable generalization performance, it is necessary to use sufficiently large and balanced training samples. To satisfy these requirements while ensuring that basic observational parameters are available for each GRB, we selected four physical parameters from the BAT sample as input features for training: spectral index ($\Gamma$, obtained from simple PL fits), peak flux ($F_{\rm p}$), fluence ($S_{\rm \gamma}$), and duration ($T_{\rm 90}$; hereafter, these four observational parameters are referred to as ``input features''). GRBs lacking any of these four parameters were excluded, resulting in the removal of 162 GRBs. The final BAT sample thus contains 1166 GRBs with complete observational data, of which only 516 are simultaneously detected by \textit{Fermi} or Konus-Wind and have reliable $E_{\rm p}$ measurements (248 observed by \textit{Fermi} and 268 observed by Konus-Wind).

The sample was ultimately divided into a training set and a generalization set. The training set was used to train the ML models, enabling them to acquire stable and reliable generalization capability, while the generalization set was employed to estimate the originally unknown $E_{\rm p}$ values using the trained models. Specifically, the training set consists of 516 GRBs with reliable $E_{\rm p}$ measurements and corresponding BAT observational parameters, including 470 long GRBs (LGRBs) and 46 short GRBs (SGRBs), which are listed in Table \ref{training_set}. The generalization set comprises 650 GRBs lacking $E_{\rm p}$ values but possessing complete observational data, including 591 LGRBs and 59 SGRBs.

\begin{table}
\caption{List of the 516 BAT GRBs in the training set (extract).}
\label{training_set}
\centering
\scriptsize
\begin{tabular}{lccccc}
\toprule
GRB & $T_{90}$ & $F_{\rm p}$ & $S_{\gamma}$ & $\Gamma$ & $E_{\rm p}$\\
 & (s) & (ph cm$^{-2}$ s$^{-1}$) & ($10^{-7}$ erg cm$^{-2}$) &  & (keV)\\
\midrule
GRB050126       &       24.80   &       0.71    $\pm$   0.17    &       8.38         $\pm$   0.80    &       1.34    &$      158     ^{+     82      }_{-    41         }$      \\
GRB050315       &       95.60   &       1.93    $\pm$   0.22    &       32.20         $\pm$   1.46    &       2.11    &$      47      ^{+     4       }_{-    5         }$      \\
GRB050318       &       32.00   &       3.16    $\pm$   0.20    &       10.80         $\pm$   0.77    &       1.90    &$      34      ^{+     8       }_{-    5         }$      \\
GRB050401       &       33.30   &       10.70   $\pm$   0.92    &       82.20         $\pm$   3.06    &       1.40    &$      105     ^{+     14      }_{-    11         }$      \\
GRB050505       &       58.90   &       1.85    $\pm$   0.31    &       24.90         $\pm$   1.79    &       1.41    &$      99      ^{+     20      }_{-    13         }$      \\
\bottomrule
\end{tabular}
\tablefoot{
\textbf{}
Only a portion of this table is shown here for guidance. The full table is available at the CDS.
}

\end{table}

\begin{figure*}[htb]
\centering
\includegraphics[angle=0,scale=0.48]{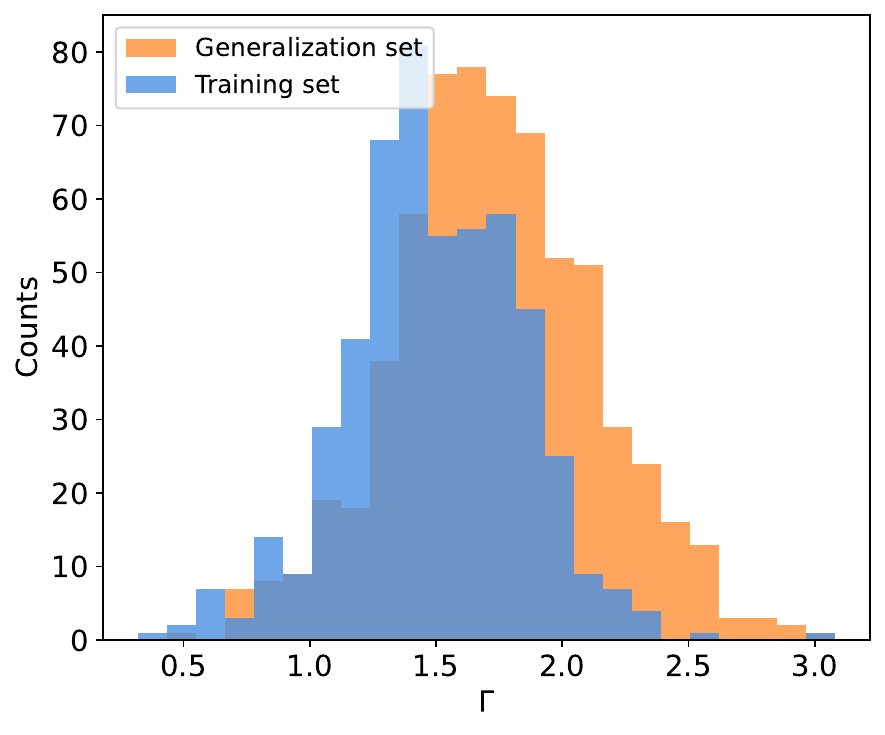}
\includegraphics[angle=0,scale=0.48]{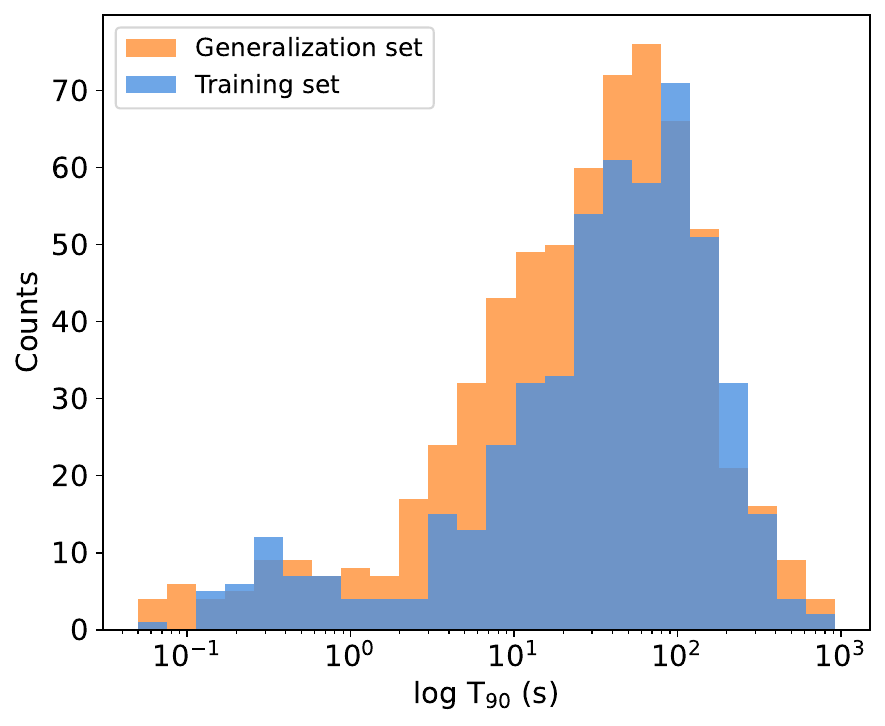}
\includegraphics[angle=0,scale=0.48]{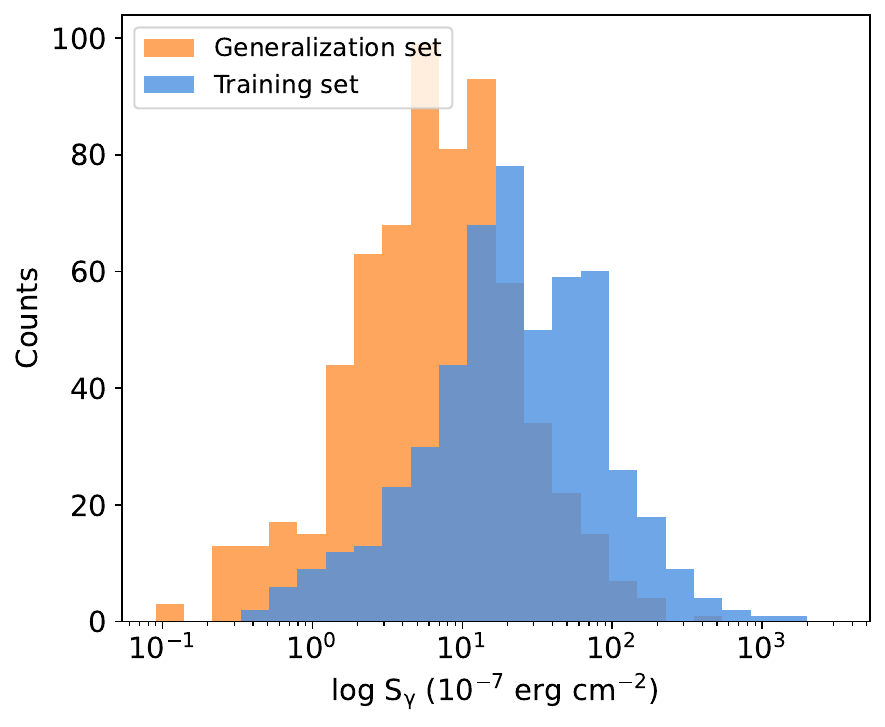}
\includegraphics[angle=0,scale=0.48]{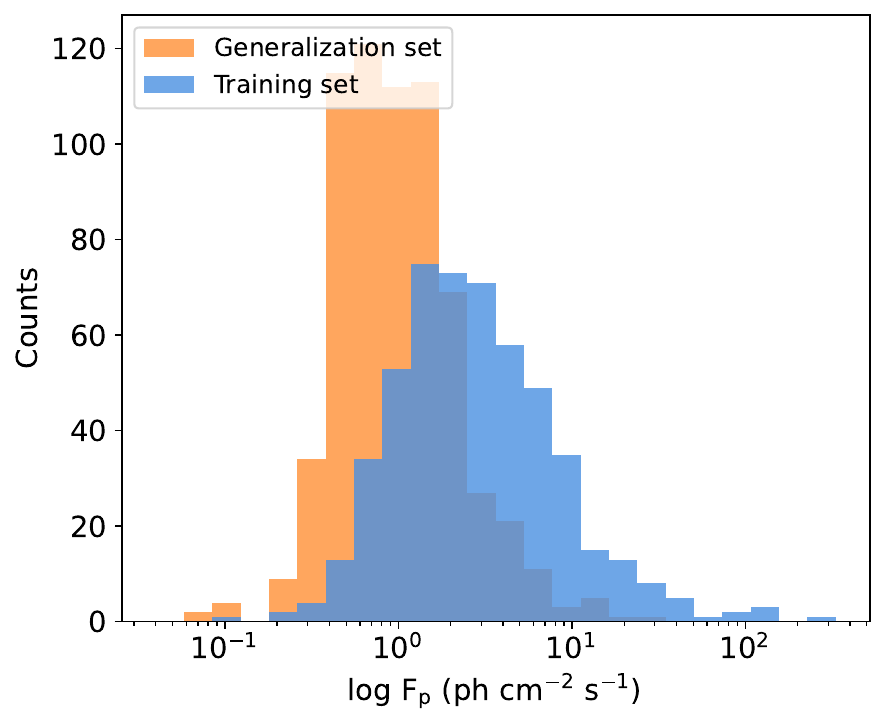}
\caption{Distributions of the four input quantities, $\Gamma$ , $T_{90}$, $S_{\rm \gamma}$, and $F_{\rm p}$, for the training set (blue region) and the generalization set (orange region).
}
\label{fig1}
\end{figure*}

\begin{figure}[ht!]
\centering
\includegraphics[width=0.45\textwidth]{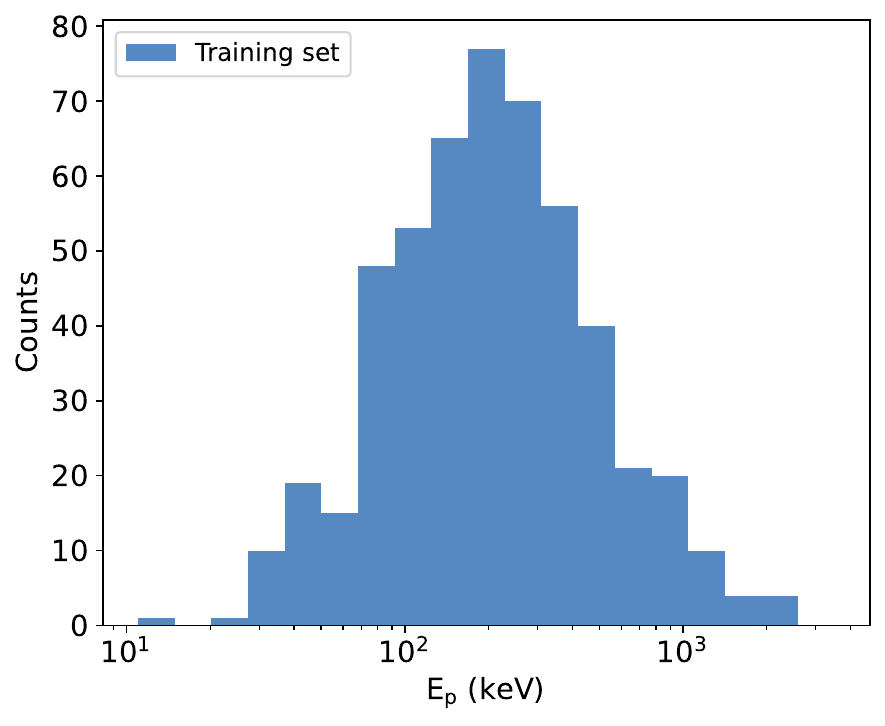}
\caption{Distribution of $E_{\rm p}$ in the training set.}
\label{fig2}
\end{figure}

It is worth noting that ML models typically require the training data to exhibit approximately normal distributions. To better satisfy this condition, all three input features except $\Gamma$ were used in logarithmic form. The parameter distributions for the training and generalization sets are displayed in Fig.~\ref{fig1}. From this figure, we find that the training set tends to have larger $F_{\rm p}$ and $S_{\rm \gamma}$ values and smaller $\Gamma$ values compared to the generalization set, while the distributions of $T_{90}$ appear broadly similar between the two sets. The distribution of $E_{\rm p}$ for the 516 GRBs in the training set is presented in Fig.~\ref{fig2}.

\begin{figure}[ht!]
\centering
\includegraphics[width=0.5\textwidth]{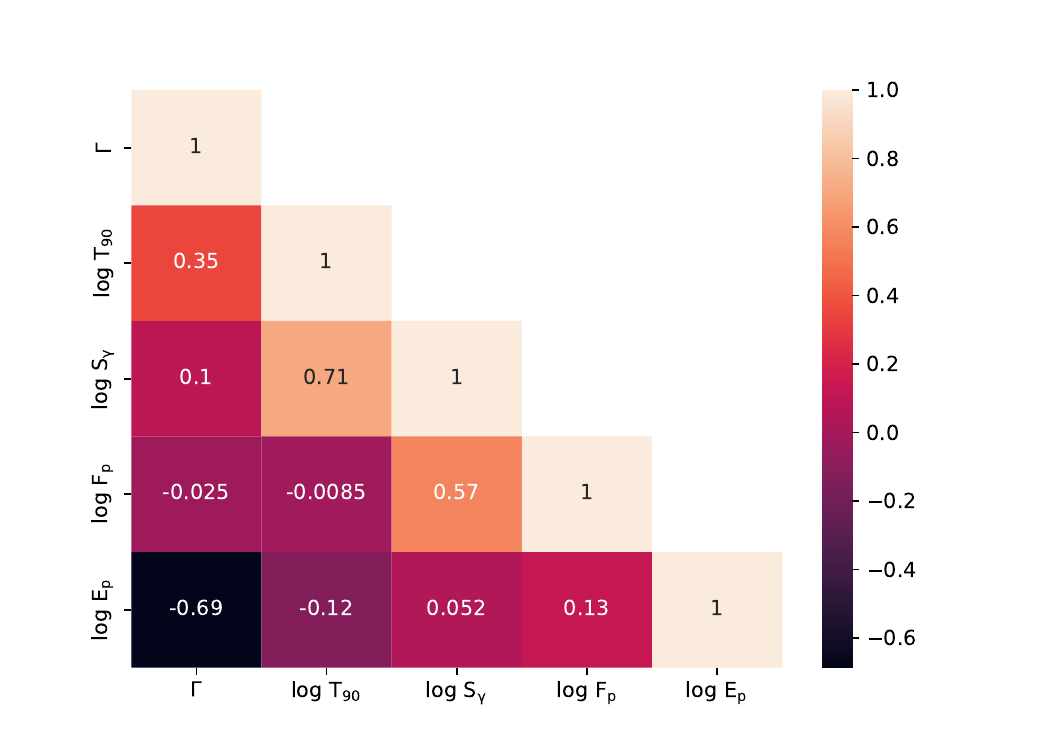}
\caption{Correlation heatmap of various parameters in the training set.}
\label{fig3}
\end{figure}

\begin{figure}[ht!]
\centering
\includegraphics[width=0.5\textwidth]{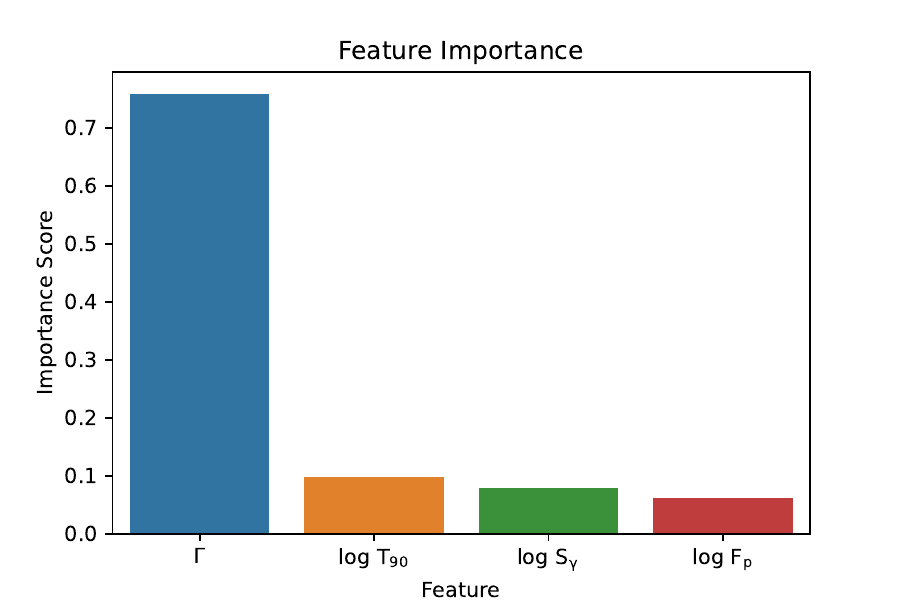}
\caption{Feature importance scores trained by the random forest algorithm, reflecting the relative importance of the four input features in the random forest training set.}
\label{fig4}
\end{figure}

In addition, to better illustrate the relationships among the parameters in the training set, we present their correlation matrix in Fig.~\ref{fig3}. The results show that $\Gamma$ exhibits the strongest negative correlation with $E_{\rm p}$ (Pearson’s $r = -0.69$), followed by $F_{\rm p}$ ($r = 0.13$) and $T_{90}$ ($r = -0.12$). Beyond linear correlation analysis, Fig.~\ref{fig4} shows the relative feature importance derived from the random forest algorithm, which reflects the contribution of each input feature to the model’s learning and estimation of $E_{\rm p}$ (see the next section for training details). The results indicate that $\Gamma$ is the most informative feature, providing the dominant contribution to the estimation of $E_{\rm p}$. In comparison, the other three parameters have weaker and relatively similar importances, with $T_{90}$ ranking second among them.

\section{Methodology} \label{sec:method}
\subsection{ML algorithms} \label{ssec:ml_model}

To obtain reliable estimates of $E_{\rm p}$ using supervised ML algorithms, we first trained the models on the training set. To maximize model accuracy and ensure stable generalization, each algorithm was trained 100 times via random data resampling: in each iteration, 80\% of the training set was used for model fitting, while the remaining 20\% served as a validation set (hereafter referred to as the test set). We evaluated model performance by comparing the estimated $E'_{\rm p}$ with the known $E_{\rm p}$ in the test set, using the root-mean-square error (RMSE) and the Pearson correlation coefficient. This procedure was repeated 100 times, and the average results were used to assess model accuracy. Notably, each of the 100 random resampling iterations involved independent model training, ensuring that the evaluation was not affected by information leakage between different subsets and thus providing an unbiased assessment of model performance.

In the second step, the trained ML models were applied to estimate $E_{\rm p}$ in the generalization set. We ultimately employed the SuperLearner technique to combine multiple trained models into a single ensemble, leveraging the strengths of each model to generate the estimated $E'_{\rm p}$ values. The individual estimations were then aggregated to produce the final estimates, which allowed the SuperLearner to achieve greater accuracy than any single algorithm alone. The ML algorithms used in this work are listed below:
\begin{enumerate}
\item 
The random forest algorithm is an ensemble method based on decision trees. It constructs multiple independent decision trees, each trained separately, and combines their outputs to improve the accuracy of the final prediction \citep{Breiman2001}. Random forest has been widely applied in both classification and regression tasks, particularly excelling in handling high-dimensional and complex data. In addition, the algorithm is sensitive to outliers and requires careful consideration when evaluating feature importance and handling missing data. An important aspect of using random forest is selecting the optimal tree depth and the number of trees according to the complexity and structure of the data.

\item
The Extreme Gradient Boosting (XGBoost) algorithm also achieves its performance by combining multiple decision trees and represents an improved version of gradient-boosted decision trees, addressing some limitations of traditional decision trees \citep{Chen2016}. It enhances model accuracy by sequentially fitting the residuals of previous predictions, with each tree trained based on the errors of the preceding tree, thereby iteratively improving predictive performance. Additionally, XGBoost incorporates regularization to control tree complexity and prevent overfitting, and employs parallelization to accelerate computation, resulting in higher efficiency and prediction accuracy.

\item
Linear regression is one of the simplest and most widely used regression models. Its core idea is to establish a linear relationship between the predicted outcome and the input features, with model parameters estimated by minimizing the sum of squared errors. 

\item
Kernel ridge regression extends linear regression into the feature space. Unlike traditional ridge regression, the kernelized version replaces the inner product with a kernel function, allowing nonlinear patterns in the original low-dimensional space to be mapped into a higher-dimensional space. A ridge regression model is then constructed in this high-dimensional space to predict new data points. The algorithm also incorporates a regularization term to prevent overfitting, thereby improving the model’s generalization ability.
\end{enumerate}

\subsection{Model training and optimization} \label{ssec:training_optimization}

During the training of supervised ML algorithms, hyperparameter tuning was performed to ensure that each algorithm achieved optimal performance on the data without underfitting or overfitting. Taking the random forest algorithm as an example, excessively deep trees can lead to overfitting and reduced generalization, whereas too few trees can result in underfitting and lower predictive accuracy. Therefore, methods such as cross-validation and grid search are typically used to determine the optimal tree depth and number of trees. Generally, increasing the number of trees improves accuracy but also increases computational complexity, requiring a balance between predictive performance and efficiency.

\begin{figure*}[htbp]
\centering
\subfloat[]{
  \includegraphics[width=0.49\textwidth]{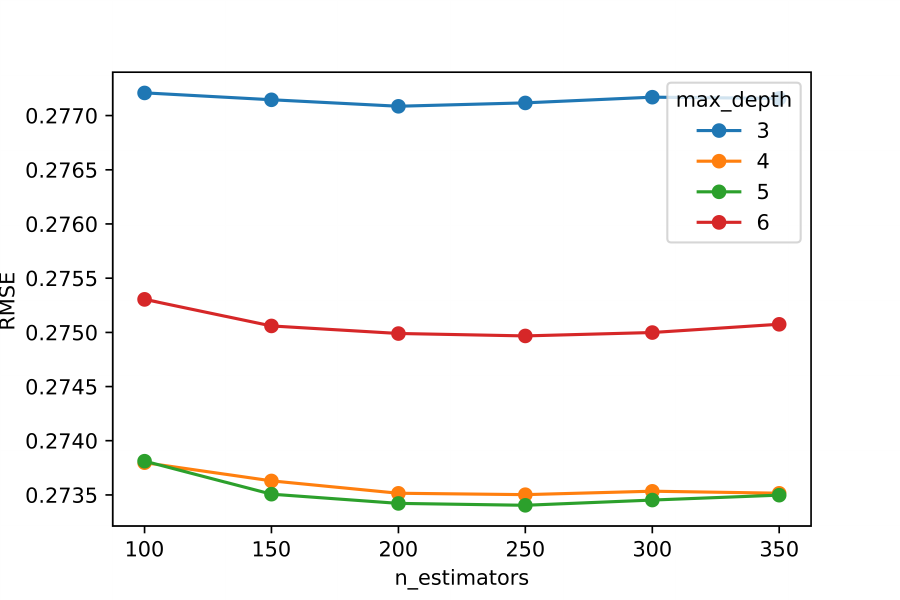}
}
\hfill
\subfloat[]{
  \includegraphics[width=0.49\textwidth]{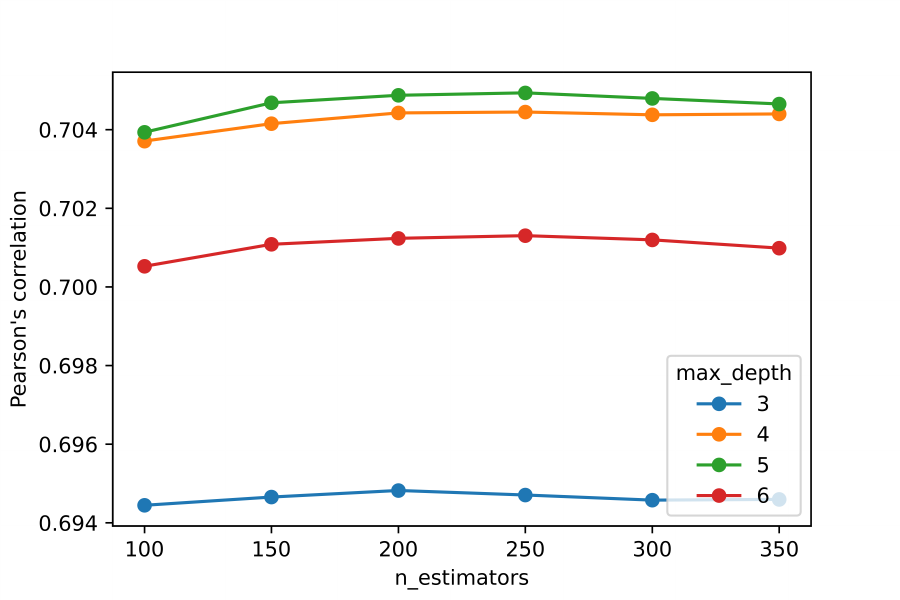}
}
\caption{
Optimization plot of hyperparameters in the random forest algorithm.
Panels (a) and (b) show the variations in RMSE and Pearson correlation coefficient with the number of trees and the maximum tree depth, respectively.}
\label{fig5}
\end{figure*}

In this work, the optimization criteria were the lowest RMSE and the highest Pearson correlation coefficient observed during tuning. The algorithms primarily optimized in this study were the tree-based ensemble methods, random forest and XGBoost. For random forest, the main hyperparameters adjusted include the number of trees and tree depth, while the maximum number of leaf nodes is kept at the default value due to the limited number of input features. The optimization process for random forest is illustrated in Fig.~\ref{fig5}, which shows the algorithm’s performance across different tree numbers and depths. The lowest RMSE and highest Pearson correlation were achieved with a tree depth of 5 and 250 trees, which were adopted as the optimal hyperparameters in this study. All other parameters were set to their default values.

\begin{figure*}[htbp]
\centering

\subfloat[]{
  \includegraphics[width=0.49\textwidth]{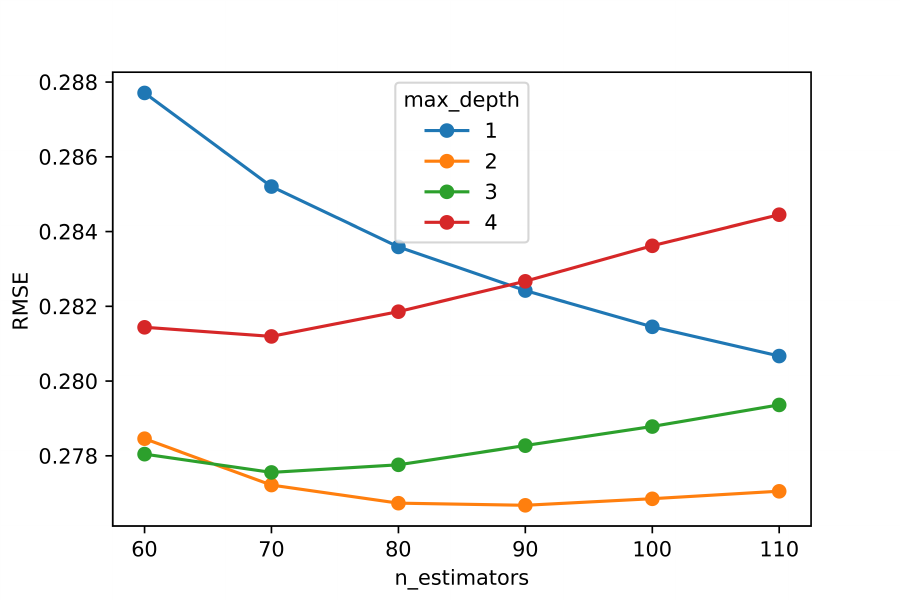}
}
\hfill
\subfloat[]{
  \includegraphics[width=0.49\textwidth]{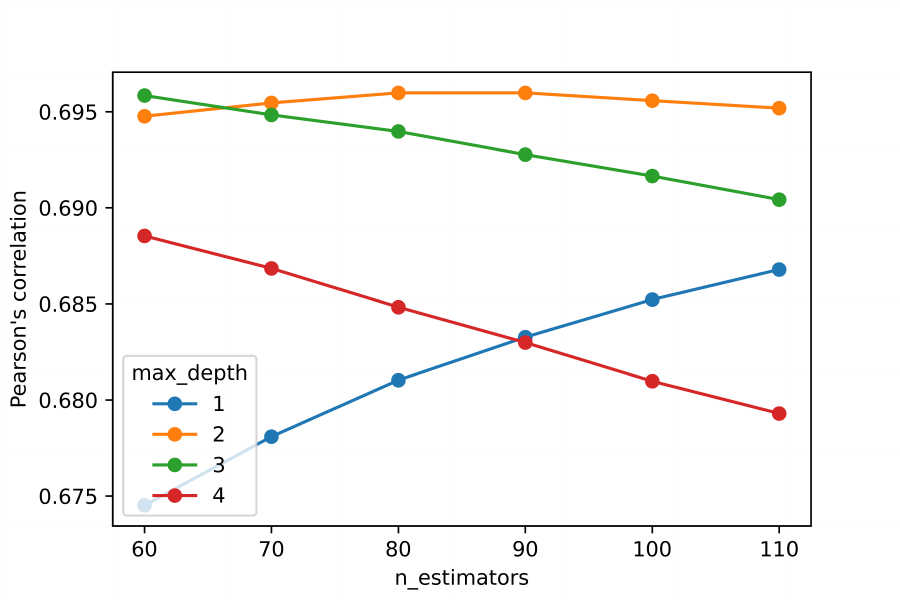}
}

\subfloat[]{
  \includegraphics[width=0.49\textwidth]{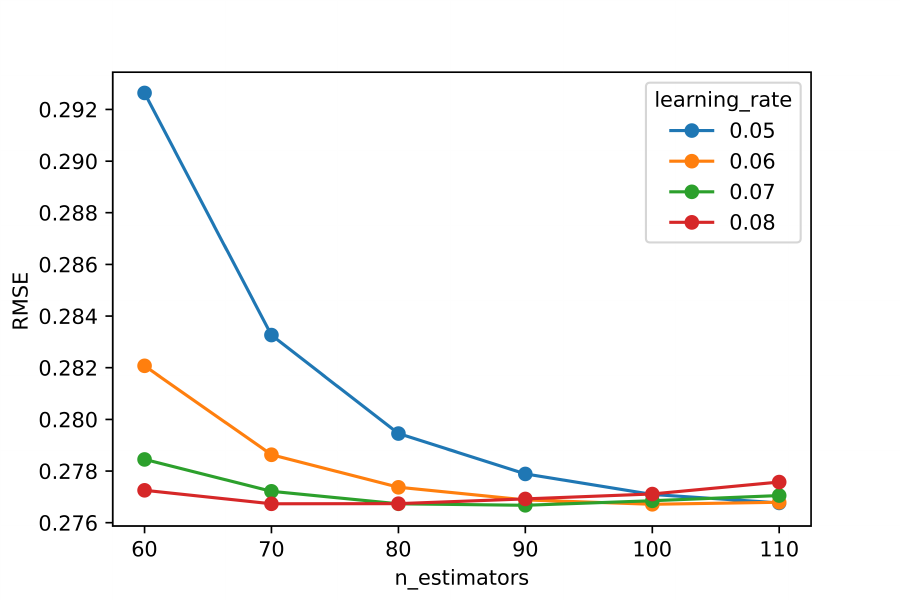}
}
\hfill
\subfloat[]{
  \includegraphics[width=0.49\textwidth]{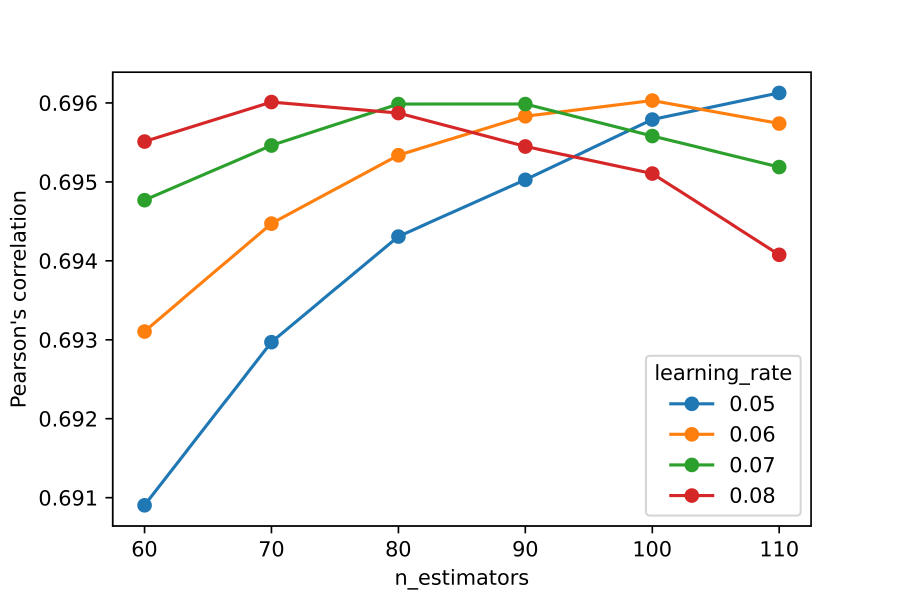}
}

\caption{
Optimization of XGBoost hyperparameters.
Panels (a) and (b) show the dependence of the RMSE and the Pearson correlation
coefficient on the number of trees and the maximum tree depth, respectively.
Panels (c) and (d) show their dependence on the number of trees and the learning
rate, respectively.
}
\label{fig6}
\end{figure*}

A similar optimization procedure was applied to the XGBoost algorithm. We first explored the dependence of model performance on the tree depth and the number of trees. As shown in Figs.~\ref{fig6}a and 6b, the residuals and Pearson correlation coefficients indicate that the optimal configuration is achieved with a tree depth of 2 and 90 trees. We then fixed the tree depth at this value and examined the impact of different learning rates for varying numbers of trees, as illustrated in Figs.~\ref{fig6}c and 6d. These results show that the best performance is obtained with 90 trees and a learning rate of 0.07. All remaining XGBoost parameters were kept at their default settings. The linear regression model used in this work contains no essential tunable hyperparameters, while the kernel ridge regression model adopted a polynomial kernel with $\alpha=0.6$, degree $=2$, and $\mathrm{coef0}=2.5$.

Different ML algorithms exhibit distinct strengths on the same dataset. In this work, we employed the SuperLearner algorithm to combine multiple models and maximize overall performance, with the detailed procedure described in the following subsection.

\subsection{SuperLearner} \label{ssec:superlearner}

After training each individual ML algorithm 100 times on randomly sampled datasets, we integrated them into a SuperLearner ensemble. SuperLearner is an ensemble method that combines multiple base algorithms using V-fold cross-validation \citep{Van2007}. For a given dataset, it is often difficult to determine a priori which algorithm will perform best, and SuperLearner effectively addresses this challenge. During cross-validation, it adaptively assigns higher weights to algorithms that are better suited to the current training data (hereafter also referred to as predictors) while reducing the weights of lower-performing models. Specifically, weights are assigned to each algorithm based on its residuals during cross-validation, with algorithms producing lower residuals receiving higher coefficients; all coefficients are non-negative and sum to one. These properties allow SuperLearner to leverage the unique strengths of different algorithms, constructing an ensemble model that effectively minimizes estimation residuals.

In this work, we employed five-fold cross-validation, in which the training set of 516 GRBs was randomly divided into five complementary subsets. SuperLearner trains on four subsets while using the remaining subset as a validation set to guide model adjustment. This process is repeated five times so that each subset serves once as the validation set. During these iterations, SuperLearner automatically assigns optimal weights to each predictor based on the data, thereby optimizing the estimation of $E_{\rm p}$ across all validation sets. We repeated this entire procedure 100 times, i.e., performing 100 iterations of five-fold cross-validation with random data splits and model training, which enabled a robust assessment of the SuperLearner ensemble’s stability and reduced dependence on any specific random data partition. Considering the limited size of the current BAT training sample and the accuracy of the observational data, this step is critical for evaluating and optimizing model performance.

Generally, using too many folds could make model training overly sensitive to local features within subsets, thereby reducing generalization, whereas too few folds limit the diversity of training samples, hindering effective learning. Therefore, we chose to use five-fold cross-validation to achieve a SuperLearner ensemble with stronger generalization, rather than ten- or three-fold. In fact, our experiments confirm that this choice yields the best performance.

All ML algorithms in this work were implemented in Python. Random forest, linear regression, and kernel ridge regression were implemented using Scikit-learn\footnote{\url{https://scikit-learn.org/stable/supervised\_learning.html}}, XGBRegressor was from XGBoost\footnote{\url{https://xgboost.ai/}}, and the SuperLearner ensemble was implemented using the Mlens package (v0.2.3)\footnote{\url{http://ml-ensemble.com/}}.

\section{Results and analysis} \label{sec:resu}

\subsection{Model performance and estimation results} \label{subsec:class}

\begin{figure}[ht!]
\centering
\includegraphics[width=0.5\textwidth]{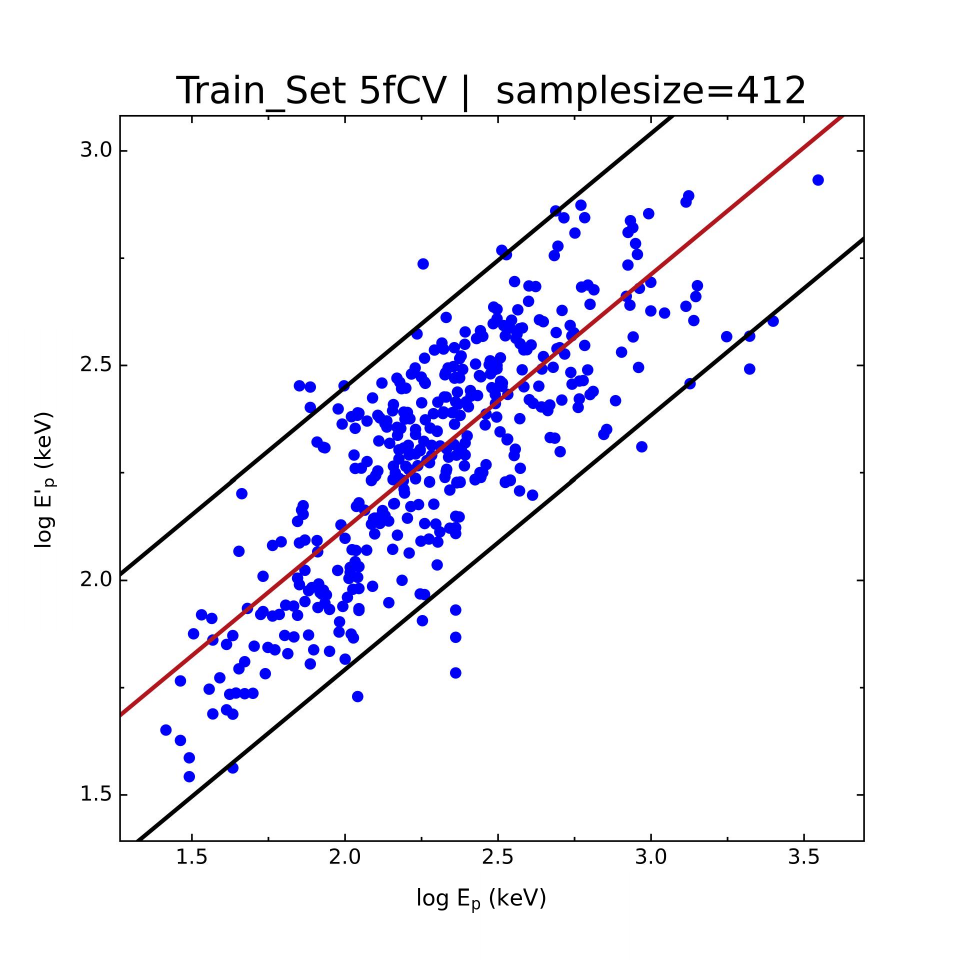}
\caption{Relation between the estimated peak energy ($E'_{\rm p}$) obtained with the SuperLearner model and the observed peak energy ($E_{\rm p}$) for the training set. The solid red line shows the best-fit relation, while the solid black lines mark the $2\sigma$ confidence bounds.}
\label{fig7}
\end{figure}

\begin{figure}[ht!]
\centering
\includegraphics[width=0.5\textwidth]{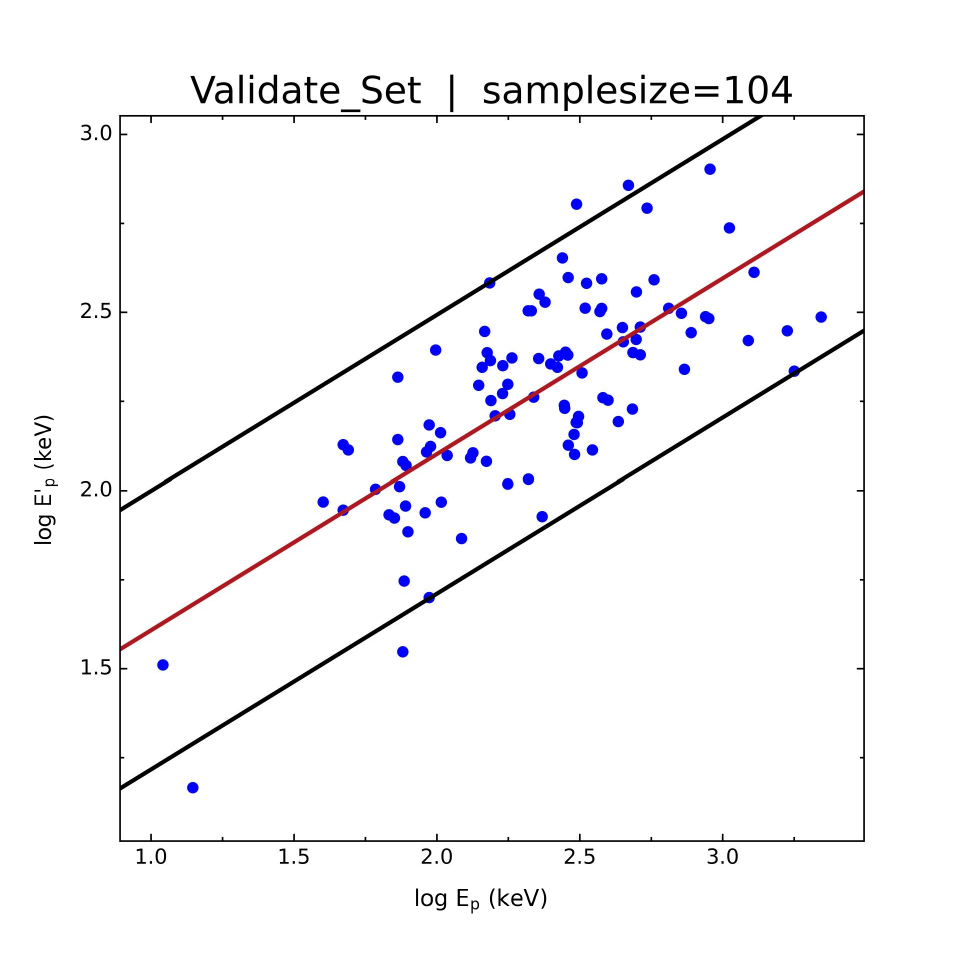}
\caption{Relation between the estimated peak energy ($E'_{\rm p}$) obtained with the SuperLearner model and the observed peak energy ($E_{\rm p}$) for the test set. The solid red line denotes the best-fit relation, and the solid black lines indicate the corresponding $2\sigma$ confidence bounds.}
\label{fig8}
\end{figure}

In this work, the final SuperLearner model used for both training and generalization integrated the four optimized predictors described in Sect. \ref{ssec:ml_model}. Figure~\ref{fig7} presents the results from 100 independent random training runs, where the horizontal axis represents the observed $E_{\rm p}$ and the vertical axis shows the model-estimated $E'_{\rm p}$. The results yield a Pearson correlation coefficient of $r = 0.725$ and a RMSE of $0.267$ between the true and estimated $E_{\rm p}$ values. According to the outlier definition proposed by \citet{2020PASP..132b4501J}, only about 5\% of the training data are identified as catastrophic outliers, i.e., GRBs with $|\Delta E_{\rm p}| > 2\sigma$, which lie outside the black lines in Fig.~\ref{fig7}. Beyond training performance, the model’s behavior on the test set is of greater significance, as it provides a more realistic measure of predictive accuracy. Figure~\ref{fig8} illustrates a representative example of the SuperLearner’s performance on the test set, where the fraction of catastrophic outliers similarly remains at approximately 5\%.

\begin{figure*}
\centering
\includegraphics[angle=0,scale=0.450]{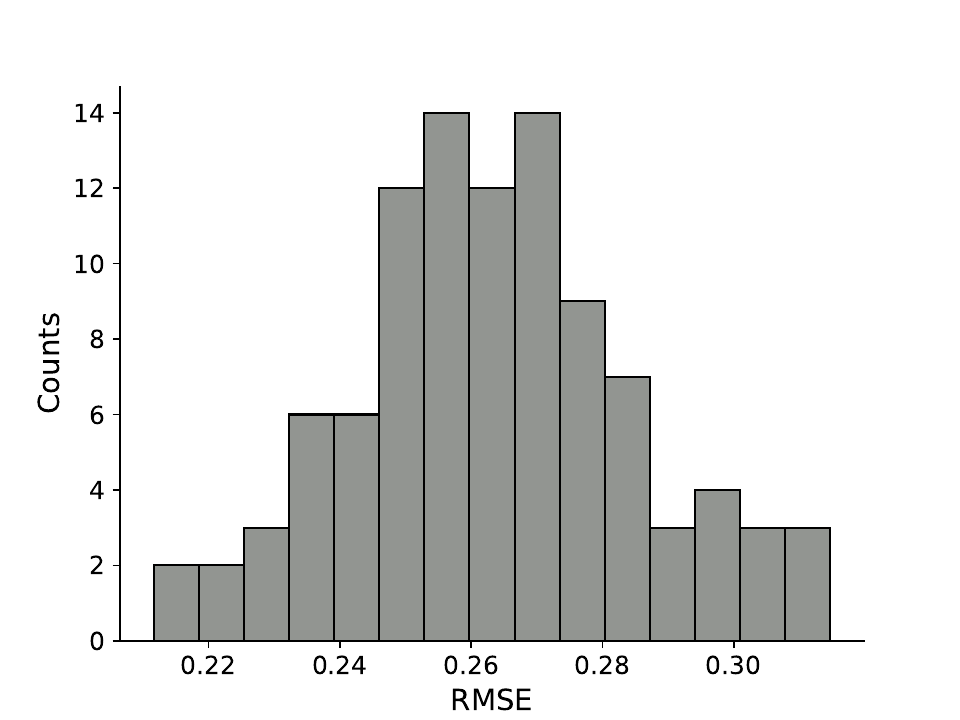}
\includegraphics[angle=0,scale=0.450]{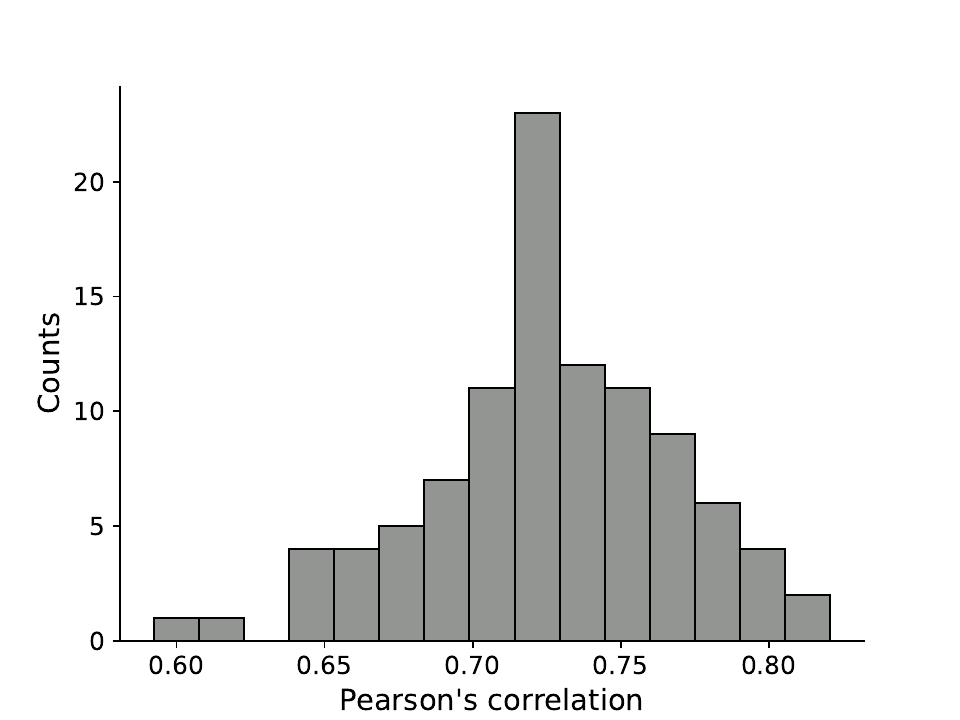}
\caption{Distributions of root mean squared error (left) and Pearson correlation coefficient (right) over 100 runs of five-fold cross-validation.}
\label{fig9}
\end{figure*}

Figure~\ref{fig9} shows the distributions of the correlation coefficients and residuals obtained from 100 rounds of five-fold cross-validation. The results from these independent training runs demonstrate that, aside from a few special cases where the predictors exhibit slightly higher or lower performance, the overall generalization ability remains stable. The averaged performance across all runs yields a Pearson correlation coefficient of $r = 0.72$ and an RMSE of $0.27$, providing a solid foundation for applying the trained model to the generalization sample.

Table \ref{performance_comparison} summarizes the performance comparison between the SuperLearner ensemble and the four individual predictors used in this study. Across 100 randomized training iterations, the SuperLearner consistently achieves higher average correlation coefficients and lower residuals, demonstrating its superior generalization performance.

\begin{table}[ht!]
\centering
\caption{Performances of SuperLearner and the four individual predictors.}
\label{performance_comparison}
\begin{tabular}{lcc}
\hline
Algorithm & $r$ & RMSE \\
\hline
SuperLearner & 0.72 & 0.267 \\
Random forest Regressor & 0.70 & 0.274 \\
XGBoost Regressor & 0.70 & 0.277 \\
Kernel Ridge & 0.71 & 0.269 \\
Linear regression & 0.70 & 0.273 \\
\hline
\end{tabular}
\end{table}

\subsection{Bias correction} \label{ssec:Bias}

It can be seen from Fig.~\ref{fig7} that some GRBs with low $E_{\rm p}$ are estimated with higher values, while those with high $E_{\rm p}$ are estimated with lower ones, a typical indication of estimation bias. Such bias usually arises from data imbalance; for example, low- or high-$E_{\rm p}$ GRBs are relatively scarce in the observed sample, preventing the model from being adequately trained on these regions. However, data imbalance is not the only cause. Limitations in the intrinsic performance of the learning algorithm or the presence of more complex nonlinear relations between input features and $E_{\rm p}$ at the low or high ends may also contribute to this issue.

We fitted a linear relation between the estimated $E'_{\rm p}$ and the observed $E_{\rm p}$ to correct for systematic bias. The correction formula is expressed as
\begin{equation}
\log E'_{\rm p}=a\times \log E_{\rm p}+b,
\end{equation}
where $E'_{\rm p}$ and $E_{\rm p}$ denote the estimated and observed peak energies, respectively, and $a$ and $b$ are the slope and intercept of the linear fit. In this work, we obtained $a = 0.606$ and $b = 0.902$. This makes it possible to correct the bias in the estimations of the generalization set, which is discussed in the following subsection.

\subsection{Estimation on the generalization set} \label{ssec:Prediction}

\begin{figure}[ht!]
\centering
\includegraphics[width=0.45\textwidth]{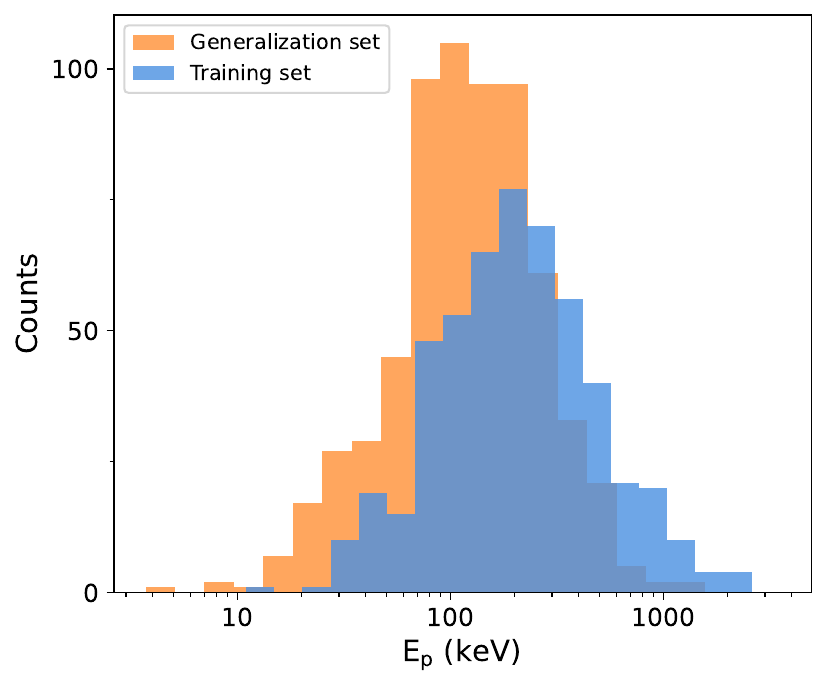}
\caption{Comparison of the peak energy distribution of the 516 GRBs in the training set (blue region) with that of the 650 GRBs in the SuperLearner generalization set (orange region).}
\label{fig10}
\end{figure}

We then applied the model to estimate the peak energies of 650 GRBs in the generalization set. The final estimation results are presented in Table \ref{generalization_set}. Figure~\ref{fig10} compares the distributions of $E'_{\rm p}$ in the generalization set and $E_{\rm p}$ in the training set. A substantial overlap is observed between the two distributions. However, the estimated $E'_{\rm p}$ values in the generalization set are systematically lower. As discussed in Sect. \ref{ssec:Fselection}, the four predictive variables in the generalization set exhibit distinct distributional differences from those in the training sample. Specifically, $F_{\rm p}$ and $S_{\rm \gamma}$ are generally lower, while $\Gamma$ tends to be higher. Considering the positive correlations of $E_{\rm p}$ with $S_{\rm \gamma}$ and $F_{\rm p}$ and its inverse correlation with $\Gamma$ (as shown in the correlation heatmap in Fig.~\ref{fig3}), the overall lower $E'_{\rm p}$ values estimated for the generalization set are therefore physically reasonable.

\begin{table*}
\caption{List of the 650 BAT GRBs in the generalization set (extract).}
\label{generalization_set}
\centering
\begin{tabular}{lcccccc}
\toprule
GRB & $T_{90}$ & $F_{\rm p}$ & $S_{\gamma}$ & $\Gamma$ & $E'_{\rm p}$ & Bias-corrected $E'_{\rm p}$ \\
 & (s) & (ph cm$^{-2}$ s$^{-1}$) & ($10^{-7}$ erg cm$^{-2}$) &  & (keV) & (keV) \\
\midrule
GRB041219C      &       4.80    &       2.45    $\pm$   0.25    &       13.10         $\pm$   0.73    &       2.02    &       57                                      &       26         \\
GRB041220       &       5.60    &       1.81    $\pm$   0.14    &       3.83         $\pm$   0.28    &       1.66    &       114                                     &       80         \\
GRB041223       &       109.10  &       7.35    $\pm$   0.32    &       167.00         $\pm$   2.79    &       1.11    &       465                                     &       820         \\
GRB041226       &       89.70   &       0.35    $\pm$   0.05    &       3.21         $\pm$   0.83    &       1.40    &       392                                     &       619         \\
GRB041228       &       55.40   &       1.61    $\pm$   0.25    &       34.90         $\pm$   1.52    &       1.60    &       168                                     &       152         \\
\bottomrule
\end{tabular}
\tablefoot{
\textbf{}
Only a portion of this table is shown here for guidance. The full table is available at the CDS. The ``Bias-corrected $E_{\rm p}$'' refers to $E'_{\rm p}$ after applying bias corrections.
}
\end{table*}

\begin{figure}[ht!]
\centering
\includegraphics[width=0.45\textwidth]{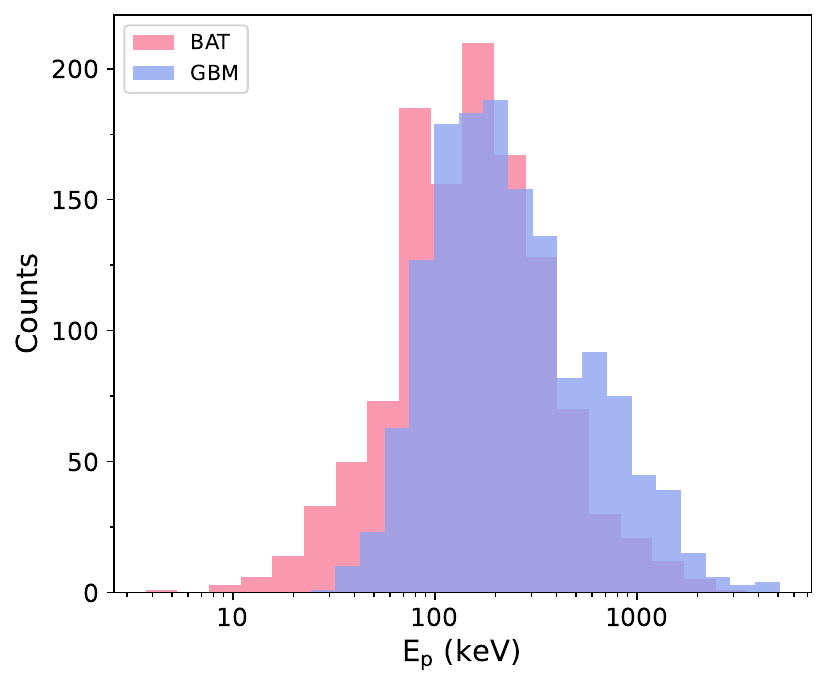}
\caption{Distribution of the peak energy of the GRBs in the BAT and GBM samples. The purple region represents the BAT sample, which is composed of the training set and the generalization set in this work. The green region represents the GBM sample, which is composed of observational data from \textit{Fermi} from between 2008 and 2018.}
\label{fig11}
\end{figure}

We further examined the overall distribution of peak energies in the BAT sample by considering the training and generalization sets as a single dataset. Observations from the past decade by the \textit{Fermi} satellite provide a useful reference \citep{2020ApJ...893...46V, 2021ApJ...913...60P}, as the GBM’s wide energy coverage and large sample size make it highly representative. Figure~\ref{fig11} compares $E_{\rm p}$ distributions of the BAT and GBM samples, showing that the BAT sample generally exhibits lower $E_{\rm p}$ values.

To more specifically characterize the differences of peak energy, we show the following statistical results: the BAT sample exhibits an $E_{\rm p}$ range of $4$-$3516\ \rm keV$ with a median value of $152\ \rm keV$, whereas the GBM sample spans $24$-$5092\ \rm keV$ with a median value of $214\ \rm keV$. Obviously, while the BAT GRBs cover a broader $E_{\rm p}$ range, their average peak energy is lower than that of the GBM GRBs. This discrepancy likely arises from the intrinsically broader energy coverage of GBM, whereas BAT’s sensitivity at lower energies may bias the measurements of predictive parameters (e.g., lower $F_{\rm p}$ and $S_{\rm \gamma}$), which in turn affects the estimated $E_{\rm p}$. Beyond instrumental effects, limitations in the ML models also contribute to residual biases. To further mitigate potential biases in the estimated $E'_{\rm p}$, we applied the correction formula described in Sect. \ref{ssec:Bias}; the corrected results are listed in Table \ref{generalization_set}. After correction, the highest estimated $E'_{\rm p}$ in the generalization set exceeds $6000\ \rm keV$, and the number of GRBs with $E'_{\rm p}>200\ \rm keV$ increases substantially, better reflecting the true distribution of GRB peak energies. The present results are based on training with only 516 GRBs, and the dataset size remains limited. Nevertheless, the proposed model provides a valuable framework for estimating the peak energies of BAT GRBs. As future joint observations from \textit{Fermi}, Konus-Wind, and \textit{Swift} accumulate more GRB data, this limitation is expected to be alleviated, enabling more reliable estimations of BAT GRB peak energies.

\subsection{Spectrum-energy correlations} \label{ssec:Spectral}

The Amati and Yonetoku relations are two of the most widely studied spectrum-energy correlations in GRBs \citep{Amati2002, 2004ApJ...609..935Y, Amati2006, 2012ApJ...750...88Z, 2019NatCo..10.1504I}. By estimating peak energies for a large \textit{Swift} sample, we examined these correlations in the BAT GRBs. We have compiled the most comprehensive redshift dataset currently available for BAT GRBs, including 392 GRBs with complete redshift and parameter measurements (Table \ref{restframe_grbs}). The redshift information was primarily taken from the website\footnote{\url{https://www.mpe.mpg.de/~jcg/grbgen.html}}. This large sample provides a valuable opportunity to revisit the spectrum-energy correlations of GRBs.

\begin{table*}
\caption{Properties of BAT GRBs with redshift (extract).}
\label{restframe_grbs}
\centering
\begin{tabular}{lccccccccc}
\toprule
GRB & $T_{90}$ & $F_{\rm p}$ & $S_{\gamma}$ & $\alpha$ & $\beta$ & $z$ & $E_{\rm p,z}$ & $\log E_{\rm iso}$ & $\log L_{\rm iso}$ \\
 & (s) & (ph cm$^{-2}$ s$^{-1}$) & ($10^{-7}$ erg cm$^{-2}$) &  &  &  & (keV) & (erg) & (erg s$^{-1}$) \\
\midrule
GRB050126       &       24.80   &       0.71    $\pm$   0.17    &       8.38         $\pm$   0.80    &       $-$1    &       $-$2.25 &       1.29    &$      362         ^{+     188     }_{-    94      }$&$    51.84   ^{+     0.04    }_{-    0.04         }$&$    51.04   ^{+     0.10    }_{-    0.10    }$      \\
GRB050315       &       95.60   &       1.93    $\pm$   0.22    &       32.20         $\pm$   1.46    &       $-$1    &       $-$2.25 &       1.949   &$      139         ^{+     12      }_{-    15      }$&$    52.82   ^{+     0.02    }_{-    0.02         }$&$    50.28   ^{+     0.05    }_{-    0.05    }$      \\
GRB050318       &       32.00   &       3.16    $\pm$   0.20    &       10.80         $\pm$   0.77    &       $-$1    &       $-$2.25 &       1.44    &$      83         ^{+     20      }_{-    12      }$&$    52.12   ^{+     0.03    }_{-    0.03         }$&$    50.18   ^{+     0.03    }_{-    0.03    }$      \\
GRB050401       &       33.30   &       10.70   $\pm$   0.92    &       82.20         $\pm$   3.06    &       $-$1    &       $-$2.25 &       2.9     &$      410         ^{+     55      }_{-    43      }$&$    53.56   ^{+     0.02    }_{-    0.02         }$&$    51.47   ^{+     0.04    }_{-    0.04    }$      \\
GRB050505       &       58.90   &       1.85    $\pm$   0.31    &       24.90         $\pm$   1.79    &       $-$1    &       $-$2.25 &       4.27    &$      522         ^{+     105     }_{-    69      }$&$    53.29   ^{+     0.03    }_{-    0.03         }$&$    51.09   ^{+     0.07    }_{-    0.07    }$      \\
GRB210420B$^{\star}$    &       158.80  &       0.50    $\pm$   0.20    &       15.00         $\pm$   2.00    &       $-$1    &       $-$2.25 &       1.4     &       467                                         &$      52.22   ^{+     0.06    }_{-    0.06         }$&$    51.04   ^{+     0.17    }_{-    0.17    }$      \\
GRB210504A$^{\star}$    &       135.06  &       0.90    $\pm$   0.20    &       27.00         $\pm$   2.00    &       $-$1    &       $-$2.25 &       2.077   &       545                                         &$      52.76   ^{+     0.03    }_{-    0.03         }$&$    51.69   ^{+     0.10    }_{-    0.10    }$      \\
GRB210517A$^{\star}$    &       3.06    &       1.50    $\pm$   0.20    &       1.50         $\pm$   0.30    &       $-$1    &       $-$2.25 &       2.486   &       334                                         &$      51.70   ^{+     0.09    }_{-    0.09         }$&$    50.45   ^{+     0.06    }_{-    0.06    }$      \\
GRB211023B$^{\star}$    &       1.30    &       2.20    $\pm$   0.30    &       1.70         $\pm$   0.30    &       $-$1    &       $-$2.25 &       0.862   &       133                                         &$      50.89   ^{+     0.08    }_{-    0.08         }$&$    49.47   ^{+     0.06    }_{-    0.06    }$      \\
GRB211024B$^{\star}$    &       603.50  &       0.90    $\pm$   0.20    &       68.00         $\pm$   3.00    &       $-$1    &       $-$2.25 &       1.1137  &       481                                         &$      52.72   ^{+     0.02    }_{-    0.02         }$&$    51.10   ^{+     0.10    }_{-    0.10    }$      \\
\bottomrule
\end{tabular}
\tablefoot{
\textbf{}
Only a portion of this table is shown here for guidance. The full table is available at the CDS. An asterisk (*) denotes GRBs in the generalization set. The columns $\alpha$ and $\beta$ represent the low-energy and high-energy photon indices, respectively. For BAT GRBs that can only be fitted with a PL model, we adopted $\alpha = -1$ and $\beta = -2.25$ of the Band function to calculate the values of $E_{\rm iso}$ and $L_{\rm iso}$. If the best-fit spectral model is CPL, only the low-energy photon index is listed.
}
\end{table*}

To derive the isotropic peak energies ($E_{\rm iso}$) and luminosities ($L_{\rm iso}$) for these 392 GRBs, we considered both the Band and CPL models. Specifically, for BAT GRBs for which the CPL model is identified as the best fit and for which the low-energy spectral index ($\alpha$) and other observational parameters are available, we used the CPL model in the calculations. For the majority of GRBs that are only fitted with PL model, we adopted the Band model to estimate, assigning typical values of $\alpha=-1$ and $\beta=-2.25$ for the low- and high-energy spectral indices, respectively \citep{2000ApJS..126...19P}.

The isotropic energy and luminosity of a GRB are derived from the following relations, with all values corrected to the $1-10000\ \rm keV$ energy band:

\begin{equation}
E_{\rm iso} = \frac{4 \pi D_{\rm L}^{2} S_{\gamma} k}{(1+z)},
\end{equation}

\begin{equation}
L_{\rm iso} = 4 \pi D_{\rm L}^{2} F_{\rm p} k,
\end{equation}
where $D_{\rm L}$ is the luminosity distance, $S_{\gamma}$ and $F_{\rm p}$ denote the time-integrated fluence and peak flux, respectively, and $k$ is the correction factor, defined as

\begin{equation}
k = \frac{\int_{1/(1+z)}^{10^{4}/(1+z)} E N(E) dE}{\int_{e_{\rm min}}^{e_{\rm max}} E N(E) dE},
\end{equation}
with $e_{\rm min}$ and $e_{\rm max}$ representing the instrument’s energy band, and $N(E)$ denoting the GRB spectral model. In this work, we adopted the Band and CPL spectral forms as in Eqs. \ref{Band} and \ref{cpl}:

\begin{equation}\label{Band}
        N(E) = 
        \begin{cases} 
                A\left( \frac{E}{100\,\mathrm{keV}} \right)^\alpha 
                \exp\left( -\frac{E}{E_0} \right), & E < (\alpha - \beta)E_0, \\[8pt]
                \begin{aligned}
                        & \textstyle A\left( \frac{E}{100\,\mathrm{keV}} \right)^\beta 
                        \left[ \frac{(\alpha - \beta)E_0}{100\,\mathrm{keV}} \right]^{\alpha-\beta} \\[4pt]
                        & \textstyle \qquad \times \exp(\beta - \alpha),
                \end{aligned} & E \geq (\alpha - \beta)E_0,
        \end{cases}
\end{equation}

\begin{equation}\label{cpl}
N(E) = A \left( \frac{E}{100~\rm keV} \right)^{\alpha} \exp \left( - \frac{E}{E_c} \right),
\end{equation}
where $\alpha$ and $\beta$ are the low- and high-energy photon spectral indices, and $E_c$ and $E_0$ denote the cutoff energies, with $E_c = E_0 = E_{\rm p}/(2+\alpha)$, where $E_{\rm p}$ is the peak energy. The final calculated values are listed in Table \ref{restframe_grbs}, with all uncertainties propagated via standard error propagation.

\begin{figure}[ht!]
\centering
\includegraphics[width=0.5\textwidth]{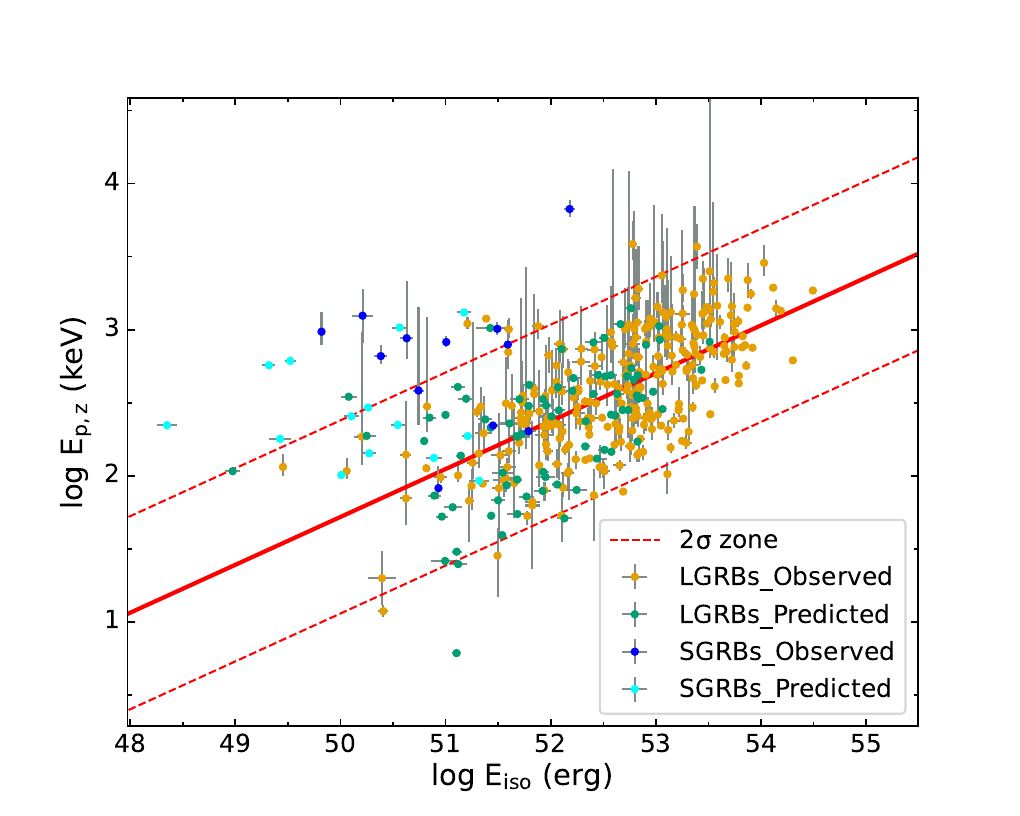}
\caption{Relationship between $E_{\rm p,z}$ and $E_{\rm iso}$ of BAT GRBs. Deep blue (light blue) dots represent SGRBs in the training (generalization) set, and orange (green) dots represent LGRBs in the training (generalization) set. The solid red line represents the best-fit line for all LGRBs, and the dashed line represents the 2$\sigma$ confidence interval.}
\label{fig12}
\end{figure}

\begin{figure}[ht!]
\centering
\includegraphics[width=0.5\textwidth]{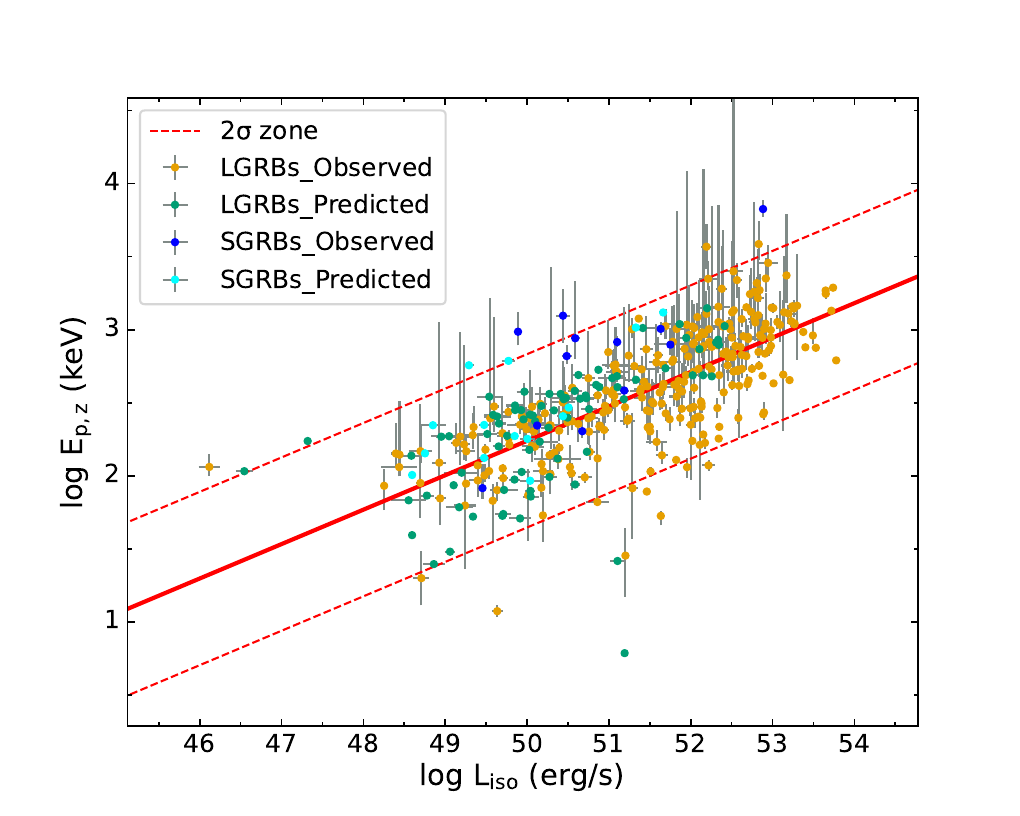}
\caption{Relationship between $E_{\rm p,z}$ and $L_{\rm iso}$ of BAT GRBs. The caption is the same as that of Fig. \ref{fig12}.}
\label{fig13}
\end{figure}

Figures~\ref{fig12} and \ref{fig13} show the Amati and Yonetoku relations for LGRBs, respectively. Orange and green circles denote GRBs with measured $E_{\rm p}$ (285 GRBs) and those with estimated $E'_{\rm p}$ from the SuperLearner model (81 GRBs). We analyzed all LGRBs together. Using a least-squares fitting, we obtained the Amati relation for the \textit{Swift}/BAT LGRB sample as
\begin{equation}
\log E_{\rm p,z} = (-14.67 \pm 1.05) + (0.33 \pm 0.02) \times \log E_{\rm iso}.
\end{equation}
This result is consistent with previous studies \citep{2020MNRAS.492.1919M,2023ApJ...950...30Z}. In our BAT sample, a clear positive correlation persists between $\log E_{\rm p,z}$ and $\log E_{\rm iso}$ for LGRBs, with a Pearson correlation coefficient of $r = 0.65$ and a slope of $a = 0.33$. We find that the 366 LGRBs in our BAT sample are consistent with the updated Amati relation reported recently by \citet{2023ApJS..266...31L}.

Similarly, analysis of the LGRB sample reveals a very tight correlation between $\log E_{\rm p,z}$ and $\log L_{\rm iso}$, as shown in Fig.~\ref{fig13}. Both the estimated and observed LGRBs follow a consistent trend, with no noticeable differences. Furthermore, we fitted the correlation between the peak energy and the isotropic luminosity for LGRBs, yielding the following Yonetoku relation:

\begin{equation}
\log E_{\rm p,z} = (-9.53 \pm 0.59) + (0.24 \pm 0.01) \times \log L_{\rm iso}.
\end{equation}

The Pearson correlation coefficient between $E_{\rm p,z}$ and $L_{\rm iso}$ is 0.73, with a slope of $a = 0.24$. This slope is somewhat lower than that reported in most previous studies \citep{2012ApJ...750...88Z,2018PASP..130e4202Z,2023ApJ...950...30Z}, although some works have also found coefficients around 0.2 \citep{2023A&A...673A..20X}. It remains uncertain whether this deviation originates from the estimation errors in $E'_{\rm p}$, from intrinsic variations in a larger sample, or from selection effects. Nevertheless, the existence of a Yonetoku relation in our LGRB sample is well supported.

In addition, the SGRBs with measured $E_{\rm p}$ and those with estimated $E'_{\rm p}$ are shown as dark and light blue circles in Figs.~\ref{fig12} and \ref{fig13}, respectively. Since only 26 SGRBs have known redshifts, the limited number of data points is insufficient to reliably constrain the spectrum-energy correlations for SGRBs. Therefore, we did not plot the fitting correlation lines for them. Nevertheless, we attempted to fit the spectrum-energy relations of SGRBs, and the corresponding results are summarized in Table \ref{regression_analysis}.

Interestingly, the fitting parameters of the Yonetoku relation are found to be very similar between LGRBs and SGRBs, indicating that their $E_{\rm p,z}$-$L_{\rm iso}$ correlations are largely consistent. This similarity suggests that LGRBs and SGRBs share similar radiation mechanisms, as also reported in several previous studies \citep{2009ApJ...703.1696Z, 2009A&A...496..585G, 2013ApJ...770...32G,2023ApJ...950...30Z}. Overall, our results provide a large \textit{Swift} sample of 392 GRBs for spectrum-energy correlation analysis, showing that the 366 LGRBs in the \textit{Swift} catalog still follow the Amati and Yonetoku correlations.

\begin{table*}[ht!]
\centering
\caption{Results of regression analysis for spectrum-energy correlations.}
\label{regression_analysis}
\begin{tabular}{lllccc}
\hline
 & Correlations & Expressions & $r$ & $p$ & $\sigma$ \\
\hline
LGRB & $E_{\rm p,z}(E_{\rm iso})$ 
     & $\log E_{\rm p,z} = (-14.67 \pm 1.05) + (0.33 \pm 0.02) \times \log E_{\rm iso}$ 
     & 0.65 & $<10^{-4}$ & 0.33 \\
     & $E_{\rm p,z}(L_{\rm iso})$ 
     & $\log E_{\rm p,z} = (-9.53 \pm 0.59) + (0.24 \pm 0.01) \times \log L_{\rm iso}$ 
     & 0.73 & $<10^{-4}$ & 0.30 \\
\hline
SGRB & $E_{\rm p,z}(E_{\rm iso})$
     & $\log E_{\rm p,z} = (-3.05 \pm 5.26) + (0.11 \pm 0.10) \times \log E_{\rm iso}$
     & 0.21 & 0.29 & 0.45 \\
     & $E_{\rm p,z}(L_{\rm iso})$
     & $\log E_{\rm p,z} = (-14.15 \pm 2.94) + (0.33 \pm 0.06) \times \log L_{\rm iso}$
     & 0.75 & $<10^{-4}$ & 0.30 \\
\hline
\end{tabular}
\tablefoot{$r$ is the correlation coefficient, $p$ is the chance probability, and $\sigma$ is the dispersion.}
\end{table*}

\section{Discussion} \label{sec:discu}

\begin{figure}[ht!]
\centering
\includegraphics[width=0.5\textwidth]{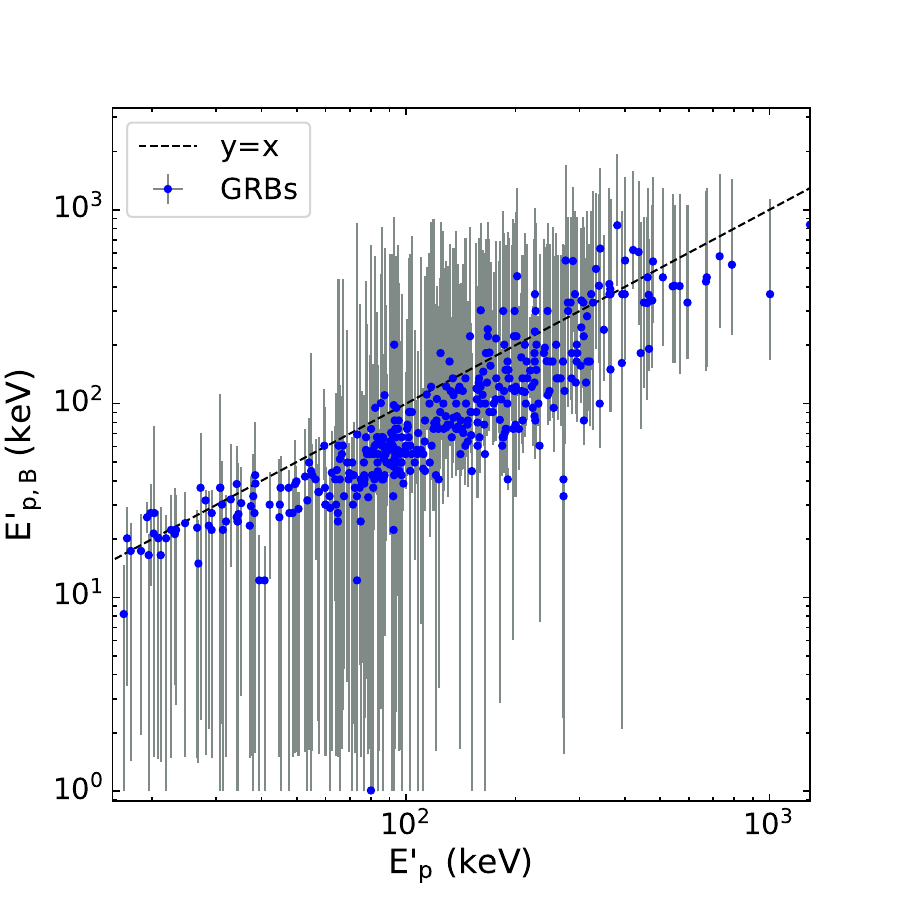}
\caption{Comparison of the estimated peak energies from the SuperLearner ($E'_{\rm p}$) and the Bayesian ($E'_{\rm p,B}$) methods.
}
\label{fig14}
\end{figure}

Since the Bayesian method for estimating GRB peak energies was proposed by \citet{2007ApJ...671..656B}, it has been widely adopted. We collected 728 estimated peak energies of GRBs from the online repository of \citet{2007ApJ...671..656B} up to December 2018. To evaluate the performance of the SuperLearner method, we selected 419 GRBs for which the estimated peak energies are available from both methods (i.e., the overlapping sample) for comparison. As shown in Fig.~\ref{fig14}, the peak energies of GRBs estimated by the Bayesian method tend to be lower than those estimated by the SuperLearner. After accounting for the 90\% credible intervals of the Bayesian estimates, the estimated peak energies from the two methods are broadly consistent for the majority of GRBs. However, for approximately 18\% (77/419) of the GRBs, the SuperLearner estimations fall outside the Bayesian credible intervals. These discrepancies largely arise because the Bayesian upper bounds are often too low to encompass the higher $E'_{\rm p}$ values inferred by the SuperLearner.

\begin{figure}[ht!]
\centering
\includegraphics[width=0.5\textwidth]{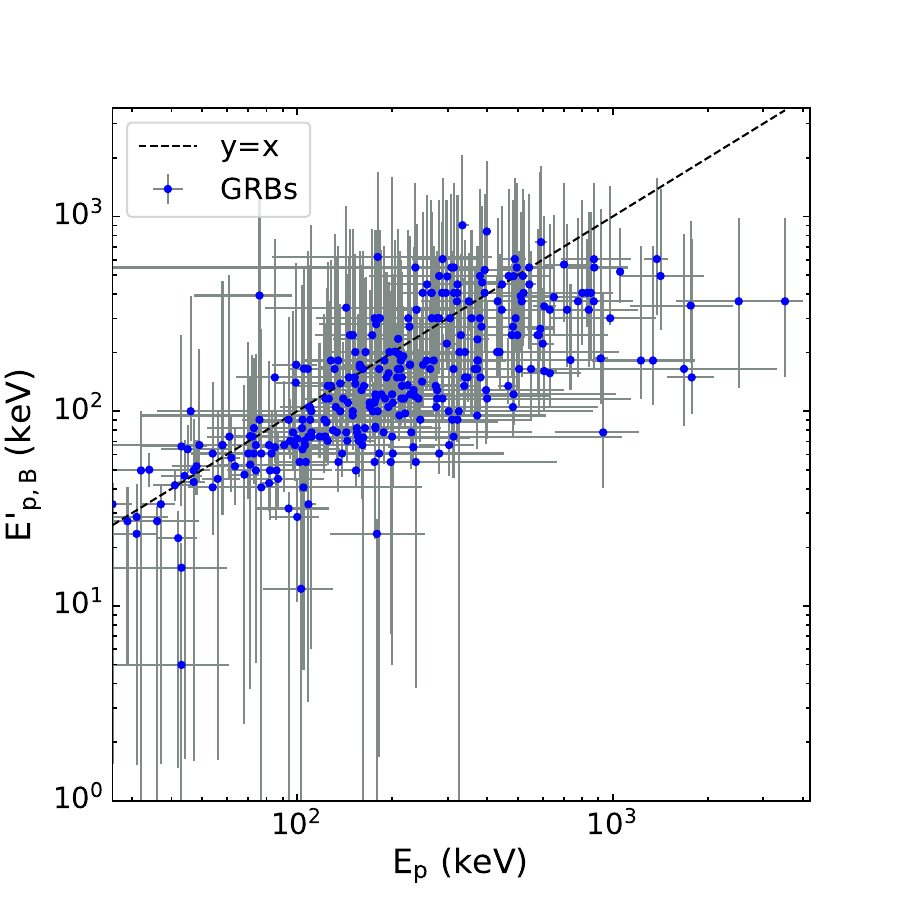}
\caption{Comparison of the observed peak energy with the estimated peak energy derived from the Bayesian method.}
\label{fig15}
\end{figure}

To more directly assess the discrepancy between estimated and observed values, we further selected 309 GRBs from 2004--2018 with both reliable observed $E_{\rm p}$ and Bayesian-derived $E'_{\rm p,B}$ (Fig.~\ref{fig15}). The horizontal axis represents the observed $E_{\rm p}$, and the vertical axis shows the Bayesian-estimated $E'_{\rm p,B}$. It is evident that the Bayesian estimates are systematically lower than the observed values, particularly for $E_{\rm p} > 60~\rm keV$. Furthermore, even when simultaneously accounting for the Bayesian 90\% credible intervals and the reported observational uncertainties of the observed $E_{\rm p}$, $\sim 14\%$ (44/309) of the GRBs show no overlap between the two ranges. Importantly, in these non-overlapping cases, the observed peak energies predominantly exceed the Bayesian upper bounds. Both the comparison with the SuperLearner estimations and the direct observational validation consistently indicate a systematic underestimation of peak energies by the Bayesian approach. Taken together, these comparisons provide indirect evidence that the SuperLearner estimations are likely less biased and closer to the intrinsic peak energies.

\begin{figure}[ht!]
\centering
\includegraphics[width=0.5\textwidth]{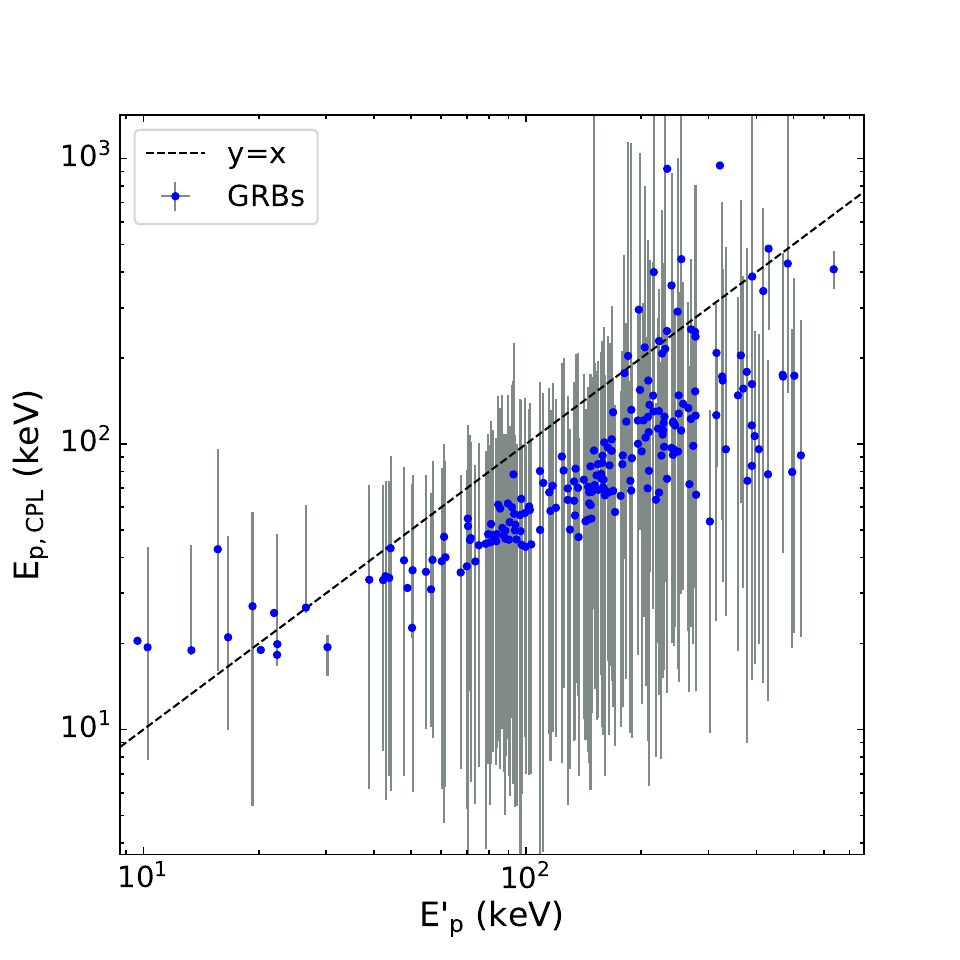}
\caption{Comparison of the peak energies estimated from the SuperLearner method with those obtained from the CPL model fitting.}
\label{fig16}
\end{figure}

To further assess the reliability of the SuperLearner model, we compiled the CPL-fitted peak energies ($E_{\rm p,CPL}$) of 229 GRBs from the BAT sample, which we obtained from the online table\footnote{\url{https://swift.gsfc.nasa.gov/results/batgrbcat/index_tables.html}}. The estimated peak energies of these GRBs were then estimated with the SuperLearner ensemble and compared with the $E_{\rm p,CPL}$. Figure~\ref{fig16} shows the relation between the SuperLearner-estimated $E'_{\rm p}$ and $E_{\rm p,CPL}$. It is evident that $E_{\rm p,CPL}$ tends to be systematically lower than $E'_{\rm p}$. This discrepancy arises because \textit{Swift}/BAT is primarily sensitive to low energies, which limits its ability to constrain the prompt radiation spectrum. As a result, direct spectral fitting of \textit{Swift} GRBs tends to systematically underestimate the peak energy \citep{2007MNRAS.382..342C}. In contrast, the SuperLearner model trained with the peak energies measured by \textit{Fermi}/GBM or Konus-Wind effectively alleviates this bias, thereby indirectly confirming the physical reliability of its estimations.

\section{Conclusions} \label{sec:concl}

We propose a SuperLearner-based framework that integrates multiple supervised ML algorithms to estimate the peak energy of \textit{Swift} GRBs using multidimensional observational parameters, i.e., $\Gamma$, $F_{\rm p}$, $S_{\gamma}$, and $T_{90}$. This method fully exploits the intrinsic correlations among these quantities, overcoming the limitations of traditional statistical methods in modeling nonlinear relationships and thereby enabling a more accurate and robust estimation of GRB peak energies. The main conclusions of this study are summarized as follows:

\begin{enumerate}
    \item Through 100 iterations of five-fold cross-validation, the SuperLearner model achieves an average RMSE of 0.27 and a correlation coefficient of $r = 0.72$ on the test set, demonstrating superior generalization performance compared to individual algorithms. 

    \item A comparison with the Bayesian estimation method proposed by \citet{2007ApJ...671..656B} shows that the SuperLearner estimations are more consistent with the true peak energies. This work therefore provides a new and more reliable approach for estimating the peak energies of \textit{Swift} GRBs.

    \item Based on this method, we estimated the peak energies for 650 GRBs in the BAT sample. To account for potential biases introduced by ML algorithms, we further performed bias correction on the estimated values, and the corrected results are listed in Table \ref{generalization_set}.

    \item Using the estimated peak energies together with 392 GRBs with measured redshifts, we revisited the $E_{\rm p,z}$-$E_{\rm iso}$ and $E_{\rm p,z}$-$L_{\rm iso}$ correlations for the \textit{Swift} sample. Our results show that both the Amati and Yonetoku relations remain evident for 366 LGRBs in the BAT sample. Although the number of SGRBs in this analysis is limited, their distribution in the Yonetoku relation exhibits a nearly identical trend to that of LGRBs, suggesting that the two classes share similar radiation mechanisms.
\end{enumerate}

\noindent This study represents the first attempt to estimate GRB peak energies by combining multidimensional observational parameters with supervised ML techniques. The proposed approach benefits from both the extensive observational data accumulated by multiple detectors and the rapid advancement of ML methodologies, offering a new perspective and a tool for GRB peak energy estimation. As more observational data become available, we will further optimize the model to improve estimation accuracy and provide more robust data support for GRB studies.  

\section*{Data availability} \label{sec:avail}
The full versions of Tables 1, 3, and 4 are available in electronic form at the CDS via \url{https://cdsarc.cds.unistra.fr/viz-bin/cat/J/A+A/708/A148}.

\begin{acknowledgements}
This work was supported in part by the National Natural Science Foundation of China (No. 12463008), and by the Guangxi Natural Science Foundation (No. 2022GXNSFDA035083).
\end{acknowledgements}

\bibliography{aa58857-26}{}
\bibliographystyle{aa}

\end{document}